\documentstyle[aps,pre,array,epsfig,eqsecnum]{revtex}

\long\def\nop#1\nopend{}

\begin{document}
\draft

\title{Can Finite Size Effects in the Poland-Scheraga Model Explain Simulations of a Simple Model for DNA Denaturation?}  
\author{Lothar Sch\"afer}
\address{Universit\"at Duisburg-Essen, Campus Essen, Universit\"atsstr.\ 5, 45117 Essen, 
        Germany}
\date{October 28, 2004}
\maketitle

\begin{abstract}
We compare results of previous simulations of a simple model of DNA denaturation to the predictions of the Poland-Scheraga model. Concentrating on the critical region of the latter model we calculate both thermodynamic quantities and the distribution functions measured in the simulations. We find that the Poland-Scheraga model yields an excellent fit to the data, provided $(i)$ we include a (singular) factor weighting the open ends of the doubly stranded chain, and $(ii)$ we keep the leading corrections to the finite size scaling limit. The exponent $c_1,$ which governs the end-weighting factor, is fairly well determined: $0.1 \alt c_1 \alt 0.15.$ The exponent $c,$ which governs the length distribution of large loops, is determined only poorly. The data are compatible with values of $c$ in at least the range $1.9 \alt c \alt 2.2.$ From the data it therefore cannot be decided whether the denaturation transition asymptotically is of first or of second order. We suggest that simulations of doubly stranded chains closed at both ends might allow for a more precise determination of $c$. 
\end{abstract} 




\section{Introduction and overview}\label{kap1}

Recent work \cite{Z1,Z2,Z3,Z4} reports on extensive simulations of a simple model for the DNA denaturation transition, i.~e. the thermal unbinding of the two strands of the DNA helix. In this model the two strands are represented by self avoiding walks of length $N$ on a (hypercubic) lattice. The walks also are mutually avoiding, but with some essential qualification. They start from the same origin and they are allowed to occupy the same lattice point, provided the distances of this point from the origin, as measured along the two strands, coincide. Such an overlap of `complementary base units' is weighted by a factor $e^{\epsilon} > 1$, where $-k_B T \epsilon$ represents a binding energy which in DNA is due to hydrogen bridges.

Clearly this model is only a rough caricature of a real DNA molecule. In particular it ignores the helical structure of the doubly-stranded parts as well as the chemical heterogeneity of DNA, which yields a base-pair dependent binding energy. However, the model fully includes the large scale properties of the embedding of the chain molecule into real space, which are governed by the excluded volume. The results of the simulations support the common view that there exists a sharp unbinding transition which in the limit of infinitely long strands, $N \to \infty,$ becomes a true phase transition. Furthermore it is suggested \cite{Z1} that this transition is of first order, with a jump in the density of bound base pairs, which is the order parameter of the transition. For some other observables, nontrivial power law scaling has been found \cite{Z1,Z2,Z3,Z4}.

On the theoretical side the standard model for the DNA denaturation transition has been proposed by Poland and Scheraga long ago \cite{Z5,Z6}. It concentrates on the internal configuration of DNA, treated as a linear sequence of doubly stranded parts and singly stranded loops and easily allows for the incorporation of the sequence heterogeneity of real DNA molecules. The Poland Scheraga (P.S.-)model has become the standard tool for analyzing denaturation experiments \cite{Z7,Z8}. Such experiments give direct access to the order parameter of the transition mentioned above. The data typically are analyzed in terms of `melting curves', defined as the derivative with respect to temperature of the average number of bound pairs. These curves may show some pronounced sequence-dependent structure, which can well be reproduced by the P.S.-model. However, the model is essentially one-dimensional. It involves only the `chemical distance' along the chain. The embedding into three-dimensional real space shows up only in some entropic weight assigned to the loops. A loop consisting of two strands of length $\ell$ is weighted by a factor $\sigma \ell^{-c},$ where the `cooperativity parameter' $\sigma$ governs the density of loops along the DNA, and the exponent $c$ is calculated from the probability that two long walks, $(\ell \gg 1),$ starting from the same point meet again after $\ell$ steps.

The numerical value of the loop exponent $c$ turns out to govern the asymptotic $(N \to \infty)$ nature of the denaturation transition. With $0 \le c \le 1$ we for small $\sigma$ in the melting curves may see some strong peak, but no proper phase transition exists. For $1 < c \le 2$ there is a second order transition, and for $c > 2$ the transition is of first order. Originally the loops were taken as closed random walks \cite{Z5}, which yields $c = d/2,$ where $d$ is the dimension of the embedding space. It immediately was pointed out \cite{Z9} that self-avoidance within the loops changes the loop exponent to $c = \nu d,$ where $\nu$ is the correlation length exponent of the self-avoiding walk problem. It takes values $\nu = 3/4 (d=2),$ or $\nu \approx 0.588 (d=3),$ respectively. More recently Kafri et.~al. \cite{Z10} invoked the theory of polymer networks to argue that the excluded volume interaction between the self-avoiding loop and the other parts of the DNA molecule increases $c$ above 2 both in two and three dimensions. For $d = 3,$ a value $c \agt 2.1$ was suggested. Furthermore, the scenario suggests that open strands of length $m,$ $1 \ll m \ll N,$ at the ends of the denaturating DNA should be weighted by a factor $m^{-c_1},$ with $c_1 \approx 0.1$ for $d =3$. Further extensions of this approach have been presented \cite{Z3}, which will be addressed below. (See Sect.~\ref{kap2.2}.)

Analyzing experimental melting curves within the framework of the P.S.-model one always takes $c_1 = 0,$ implying that no nontrivial end effects exist. The loop exponent typically is choosen as $c \approx 1.75,$ close to $c = \nu d,$ $d=3$. The success of the analysis, --though most remarkable, -- cannot be taken as experimental support of this choice, since the melting curves are not sufficient to unambiguously fix the large number of parameters involved in applying the model to a real DNA molecule \cite{Z6,Z7,Z8}. It recently has been shown \cite{Z11} that melting curves for a given base sequence calculated within the P.S.-model both with $c = 1.75$ or $c = 2.15$ can be brought on top of each other by adjusting the cooperativity parameter $\sigma$. For $c = 2.15,$ $\sigma$ has to be increased by a factor of order 10 compared to its value for $c = 1.75$. Since independent information on $\sigma$ is missing, the experiments leave the value of $c$ and thus the asymptotic character of the transition undetermined.

This situation asks for a closer examination of the results of the simulations described above. Clearly, if the P.S.-model accurately captures the physics of DNA denaturation, then it has to reproduce quantitatively the results of the simple lattice model, where the complicated chemical microstructure associated with many fit parameters is absent. The present work therefore tries to answer three related questions.\\
$(i)$ Is a model of P.S.-type able to consistently fit all the relevant simulation results?\\
$(ii)$ Does the analysis fix the exponents $c,$ $c_1$?\\
$(iii)$ If the data are compatible with a range of exponents, what other observables able to fix the exponents are in reach of present day computer experiments?

The first question is not trivially answered. In view of the experimental situation described above we clearly expect the theory to provide a good fit to the simulated melting curves, but the computer experiments provide much more detailed information. In particular, the distribution functions of both the number of bound pairs and the length of the loops have been measured. It is thus not self-evident that we find a positive answer to question $(i):$ if we include some nontrivial weighting of the open ends the thus generalized P.S.-model reproduces all the published data curves remarkably well. We stress that we here refer to the whole functions, not just to asymptotic power laws. We will find that the chain length in the simulations by far is too small to reach asymptotic power law scaling. The answer to question $(ii)$ is somewhat ambiguous. The analysis clearly shows the need of an end-weighting exponent $0.10 \alt c_1 \alt 0.15,$ but all observables related to the distributions of the number of bound pairs or the length of the open ends reasonably well can be fitted with loop exponents in the range $1.7 \alt c \alt 2.3$. It is only the measured loop length distribution which in the region of smaller $\ell$ clearly favors a value $c > 2$. Since, however, the loop weight $\sigma \ell^{-c}$ by construction is meant to hold for $\ell \gg 1,$ fixing a precise value of $c$  is somewhat a matter of taste. Insisting on fitting the data down to $\ell \approx 10,$ a value of $c = 2.05$ can be extracted. However, assuming that corrections to the asymptotic loop weight die out only for $\ell \agt 100,$ (we will give some arguments supporting this view), we can fit the data with $c$ in the range $1.9 \alt c \alt 2.2$.

In view of these results, consideration of question $(iii)$ is of interest. The trivial answer that longer chains must be simulated seems to be unrealistic. Changing the boundary conditions should be a more realistic option. In the available simulations the two strands are bound together at one end, whereas the other end of the `DNA chain' may be, and in general will be, open. Comparing to results for chains bound together of both ends may considerably restrict the allowed range of values of $c$.

With the restricted chain lengths of the simulations, (Ref.~\cite{Z1}: $500 \le N \le 3000;$ Refs.~\cite{Z2,Z3,Z4}: $80 \le N \le 1280$), our analysis necessarily concentrates on finite size effects in the P.S.-model. However, the normal finite size scaling limit is not sufficient. This limit consists in taking $N$ to infinity, with other variables, properly scaled by powers of $N$, held fixed so as to magnify the critical region. In our analysis we need to keep the leading correction terms to this limit, which for $N \to \infty$ vanish as $N^{-|c-2|}$. Since in any case $|c-2|$ is small, for the chain lengths considered these corrections are not negligible. Indeed, they explain the deviations of the effective exponents previously extracted from the data \cite{Z1,Z3,Z4} from the exponents $c,$ $c_1$ introduced in the model.

This paper is organized as follows. In Sect.~\ref{kap2} we define our version of the P.S.-model, and we derive expressions for the observables of interest. Sect.~\ref{kap3} is devoted to the finite size scaling limit. In Sect.~\ref{kap4} we take up question $(i),$ showing that with exponents $c = 2.05,$ $c_1 = 0.13$ and keeping the leading corrections to the finite size scaling limit a good fit of the simulation data is found. In Sect.~\ref{kap5} we consider the range of exponents consistent with the data, (question $(ii)$), and other observables sensitive to $c,$ (question $(iii)$), are discussed in Sect.~\ref{kap6}. Sect.~\ref{kap7} summarizes our conclusions. 

\newpage

\section{Observables of the Poland Scheraga model}\label{kap2}

The P.S.-Model describes the internal configuration of DNA as an alternating sequence of doubly stranded parts, (lengths $j_\mu$), and loops, (single strand lengths $\ell_\mu$). The weight of such a configuration is written as a product of weights ${\cal V}\left(j_\mu\right)$ for the doubly stranded parts and weights ${\cal U} \left(\ell_\mu \right)$ for the loops. A detailed analysis of the dependence  of the phase transition on the asymptotic behavior of ${\cal V}\left(j\right)$, ${\cal U}\left(\ell\right)$ recently has been presented in Ref.~\cite{Z12}. We here for ${\cal V} \left(j\right)$ take the original choice of Poland and Scheraga:
\begin{equation}
{\cal V}\left(j\right)= w^j~.
\end{equation}

The parameter $w$ absorbs both energetic and entropic effects. ${\cal U}(\ell)$ is taken to be of long range and will be specified below.

Finite size effects in one dimensional systems with long range interactions are sensitive to the boundary conditions. For the present problem two types of boundary conditions are to be considered, which in the sequel will be distinguished by indices $(bb)$ or $(bu)$, respectively. With $(bb)$-boundary conditions the two strands are bound together at both ends of the (doubly stranded) chain, whereas $(bu)$-boundary conditions allow for unbinding of the two strands at one end. 

In the first part of this Section we derive general expressions for the partition function and related quantities. For our specific choice of ${\cal U} \left(\ell \right)$ and of a factor $p\left(m, N \right)$ weighting open ends of length $m$ more explicit expressions are presented in the second part. The third part of this section is devoted to some distribution functions that have been measured.

\subsection{General structure of the partition function and related quantities}\label{kap2.1}

We first consider $(bb)$-boundary conditions. For strands of total length $N$ the partition function reads

\begin{eqnarray}
{\cal Z}_{bb}\left(N\right) = \sum^{\infty}_{k = 0} \:\: & &\sum_{\{j_0, j_1, \cdots ,j_k\} \ge 1 \atop \{\ell_1, \cdots ,\ell_k\} \ge 1} \delta \Big(N - j_0 - \sum^{k}_{\mu = 1}\left(j_\mu + \ell_\mu\right)\Big) w^{j_0} \prod^{k}_{\mu = 1} \left({\cal U}(\ell_\mu) w^{j_\mu}\right)~.
\end{eqnarray}
Here $\delta\left(n\right)$ stands for Kronecker's symbol $\delta_{n,0.}$ We define generating functions
\begin{eqnarray}
G\left(\lambda\right) & = & \sum^{\infty}_{N=0} \lambda^{-N} {\cal Z}_{bb}^{\left(N\right)} ~, \\ 
{\tilde{\cal U}} \left(\lambda\right) & = & \sum^{\infty}_{\ell=1} \lambda^{-\ell} {\cal U} \left(\ell\right) ~.
\end{eqnarray}
The explicit form (2.2) of ${\cal Z}_{bb}\left(N\right)$ results in 
\begin{equation}
G\left(\lambda\right) = \left[\frac{\lambda}{w}-1-\tilde{\cal U} \left(\lambda\right) \right]^{-1}~.
\end{equation}
Analysis of $G\left(\lambda\right)$ is sufficient to exhibit the asymptotic behavior, $\left(N \to \infty\right),$ of the model. Being interested in finite $N$, we have to invert the transform (2.3).
\begin{equation}
{\cal Z}_{bb}\left(N\right) = \oint \frac{d\lambda}{2 \pi i\lambda} \lambda^{N} G \left(\lambda\right)
\end{equation}
The integration extends over a closed path encircling all singularities of $G \left(\lambda\right)$ in the complex $\lambda$-plane.

The general ansatz for ${\cal U}\left(\ell \right)$ reads
\begin{equation}
{\cal U}\left(\ell\right) = \sigma\left(\ell^{-c} + {\cal U}_s\left(\ell\right)\right) ~,
\end{equation}
where ${\cal U}_s\left(\ell\right)$ decreases faster than $\ell^{-c}$ and corrects the asymptotic power law for smaller $\ell.$ It is easily checked that the leading term $\sigma \ell^{-c}$ in the generating function $\tilde{\cal U}\left(\lambda\right)$ gives rise to cut extending from $\lambda = 0$ to $\lambda^* = 1.$ (See Subsect.~\ref{kap2.2} for the explicit analysis.) $G\left( \lambda\right)$ inherits the cut and may show poles in addition. Specifically, for $w$ sufficiently large there exists a pole at $\lambda_1 > 1.$ For $c > 1$ this pole merges with the branch point $\lambda^* = 1$ at a critical value $w^*$ given by
\begin{equation}
\frac{1}{w^*}-1- \tilde{\cal U}\left(1\right)=0~.
\end{equation}
This equation locates the critical point of the phase transition. Exploiting the positivity of ${\cal U}(\ell)$ it is easily checked that all other poles $\lambda_p$ of $G (\lambda)$ obey $|\lambda_p| < \max$ $\left(1,\lambda_1\right)$ and therefore contribute negligibly to the behavior of long chains.

In our analysis we will restrict ourselves to the critical region, defined by $|w - w^*| = {\cal O} \left(N^{-\phi} \right),$
\begin{equation}
\phi = \min \left(1, c-1 \right)~.
\end{equation}
Up to terms of order $1/N,$ which we neglect, it then is found that all $\lambda$-integrals are dominated by the neighborhood of the branchpoint $\lambda^* = 1:$ $|\lambda - 1| = {\cal O} \left(1/N\right).$ In the sequel we will use this information in simplifying the general expressions. The partition function immediately yields the density $\rho_{bb}$ of bound pairs:
\begin{eqnarray}
\rho_{bb} & = & \frac{w}{N} \frac{\partial}{\partial w} \ln {\cal Z}_{bb} = \frac{1}{N} \: \frac{{\cal Z}^{[1]}_{bb}}{{\cal Z}_{bb}}~, \\
{\cal Z}^{[1]}_{bb} \left(N \right) & = & \frac{1}{w^*} \oint \frac{d\lambda}{2 \pi i \lambda} \lambda^{N} G^2\left(\lambda\right)~.
\end{eqnarray}
Being interested only in the critical region we in Eq.~(2.11) replaced a factor $\lambda/w$ by $1/w^*.$ The fluctuation in the number of bound pairs
\begin{equation}
C_{bb} = \frac{1}{N} \left(w \frac{\partial}{\partial w}\right)^2 \ln {\cal Z}_{bb} = w \frac{\partial}{\partial w} \rho_{bb}
\end{equation}
for shortness will be addressed as the melting curve. In the critical region it is related to the experimental melting curve via a factor relating $w - w^*$ to the deviation of the temperature from its critical value.

Another quantity of interest is the density of loops. It is defined as
\begin{equation}
\rho^{[L]}_{bb} = \frac{1}{N} \: \frac{{\cal Z}^{[L]}_{bb}}{{\cal Z}_{bb}}~,
\end{equation}
where
\begin{eqnarray}
{\cal Z}^{[L]}_{bb}\left(N\right) = \sum^{\infty}_{k = 1} k \:\: \sum_{\{j_0, j_1, \cdots ,j_k\} \ge 1 \atop \{\ell_1, \cdots ,\ell_k\} \ge 1} \delta \Big(N - j_0 - \sum^{k}_{\mu = 1}\left(j_\mu + \ell_\mu\right)\Big) w^{j_0} \prod^{k}_{\mu = 1} \left({\cal U}\left(\ell_\mu\right) w^{j_\mu}\right)~.
\end{eqnarray}
In terms of the generating functions ${\cal Z}^{[L]}_{bb}$ takes the form
\begin{equation}
{\cal Z}^{[L]}_{bb} \left(N\right) = \oint \: \frac{d\lambda}{2 \pi i \lambda} \lambda^N \tilde{\cal U} \left(\lambda\right) G^2\left(\lambda\right)~.
\end{equation}
Since in the critical region $\lambda = 1 + {\cal O} \left(1/N\right),$ we can replace the factor $\tilde{\cal U}\left(\lambda\right)$ by $\tilde{\cal U}\left(1\right) = 1/w^* -1,$ (cf. Eq.~(2.8)) to find
\begin{equation}
{\cal Z}^{[L]}_{bb}\left(N\right) = \left(1 - w^*\right) {\cal Z}^{\left[1\right]}_{bb}\left(N\right)~.
\end{equation}
We thus have the relation
\begin{equation}
\rho^{[L]}_{bb} = \left(1 - w^*\right) \rho_{bb}~,
\end{equation}
correct up to terms of order $N^{-\phi}.$

This simple result has some interesting consequences. It suggests a method for a precise determination of the critical point $w^*$ and furthermore allows us to identify the critical region as that range where the ratio $\rho_{bb}/\rho^{[L]}_{bb}$ essentially stays constant. This ratio has a simple interpretation. Up to a $1/N$ correction it is the mean length of the doubly stranded parts connecting the loops. Eq.~(2.17) thus implies that during the denaturation transition not the average length but the average number of the doubly stranded parts decreases. Essentially pairs of loops combine to form larger loops. The mean length of the loops takes the simple form 
\begin{equation}
\left<\ell\right>_{bb} = \left(\rho^{[L]}_{bb} \right)^{-1} \left(1 - \rho_{bb}\right) = \frac{1}{1 - w^*} \left(\frac{1}{\rho_{bb}}-1 \right)~.
\end{equation}

$\left<\ell\right>_{bb},$ as given by Eq.~(2.18), is the mean length averaged over all loops on the chain. A priory it differs from the mean length of a specific loop, e.g. the first one. The latter is defined as
\begin{eqnarray}
\left<\ell_1\right>_{bb} & = & \frac{{\cal Z}^{[\ell_1]}_{bb}}{{\cal Z}_{bb}}~, \\
\nonumber {\cal Z}^{[\ell_1]}_{bb}\left(N\right) & = & \sum^{\infty}_{k = 1} \:\: \sum_{\{j_0, j_1, \cdots ,j_k\} \ge 1 \atop \{\ell_1, \cdots ,\ell_k\} \ge 1} \delta \Big(N - j_0 - \sum^{k}_{\mu = 1}\left(j_\mu + \ell_\mu\right)\Big) w^{j_0} \ell_1 \prod^{k}_{\mu = 1} \left({\cal U}\left(\ell_\mu\right) w^{j_\mu}\right) \\ 
& = & \frac{w^*}{1-w^*} \oint \frac{d\lambda}{2 \pi i \lambda} \lambda^N G\left(\lambda\right) \left(- \lambda \frac{d}{d\lambda}\tilde{\cal U} \left(\lambda\right)\right)~.
\end{eqnarray}
In the last line we again replaced some term $\lambda/w$ by $1/w^*.$ We further note that in principle the definition of $\left<\ell_1\right>_{bb}$ involves the probability that at least one loop exists on the chain. However, this probability differs from $1$ only by negligible terms. In contrast to $\left<\ell\right>_{bb},$ $\left<\ell_1\right>_{bb}$ is not simply related to $\rho_{bb}.$ Physically the difference arises from the fact that $\left<\ell\right>$ gives stronger weight to short loops, which are numerous on the chain. We will find that measuring $\left<\ell_1\right>$ would allow for a more precise determination of the loop exponent $c.$

All the above results easily are transferred to $(bu)$-boundary conditions. The partition function takes the form
\begin{equation}
{\cal Z}_{bu} \left(N\right) = \sum\limits^{N-1}_{m=1} p \left(m,N \right) {\cal Z}_{bb} \left( N-m\right)~,
\end{equation}
where the factor $p\left(m,N \right)$ weights open ends of length $m.$ We recall that such a weight is absent in the original formulation of the P.S.-model. The generalization to $p \left(m, N\right) \not\equiv 1$ will turn out to be quite important. We further note that we omitted the closed configuration $m=0,$ which carries negligible weight. Quantities ${\cal Z}^{[1]}_{bu}$, ${\cal Z}^{[L]}_{bu}$, ${\cal Z}^{[\ell_1]}_{bu}$ are related to their $(bb)$-counterparts in complete analogy to Eq.~(2.21). In the critical region the relation among the density of loops and the density of bound pairs, (cf.~Eq.(2.17)), is preserved:
\begin{eqnarray*}
\rho^{[L]}_{bu} = \left(1-w^*\right) \rho_{bu}~.
\end{eqnarray*}
The only, but quite trivial, change concerns the mean length of a loop, averaged over all loops. It takes the form
\begin{equation}
\left<\ell \right>_{bu} = \frac{1}{1-w^*} \left(\frac{1-\left<m/N\right>}{\rho_{bu}} -1\right)~,
\end{equation}
where $\left<m/N \right>$ is the mean length of an open end:
\begin{equation}
\left<\frac{m}{N} \right> = \frac{1}{{\cal Z}_{bu}\left(N\right)} \: \sum\limits^{N -1}_{m = 1} m \: p\left(m,N\right) {\cal Z}_{bb} \left(N - m\right)~.
\end{equation}

\subsection{Explicit expressions valid in the critical region}\label{kap2.2}

The form (2.7) of ${\cal U}\left(\ell\right)$ yields the generating function
\begin{equation}
\tilde{\cal U}\left(\lambda\right) = \sigma \left({\cal L}i_c \left(\frac{1}{\lambda}\right) + \tilde{\cal U}_s\left(\lambda\right)\right) ~,
\end{equation}
where
\begin{equation}
{\cal L}i_c\left(z\right) = \frac{1}{\Gamma\left(c\right)} \int\limits^1_0dy \frac{\left(-\ln y\right)^{c-1}}{1/z - y}
\end{equation}
is the polylogarithmic function and $\Gamma\left(c\right)$ denotes the Gamma function. Clearly ${\cal L}i_c\left(1/\lambda\right)$ has a cut for $0 \le \lambda \le 1,$ but no other singularities. Expressions like Eq.~(2.6) show that the contribution of the cut to the $\lambda$-integral for long chains is dominated by the region $1 - \lambda = {\cal O} \left(\frac{1}{N}\right).$ In the critical region also the relevant pole $\lambda_1$ approaches $1.$ We thus write
\begin{equation}
\lambda = 1 + \frac{x}{N}~,
\end{equation}
and we expand $\tilde{\cal U}\left(\lambda\right)$ in powers of $x/N.$ Since the exponent $c$ is expected to be close to $c=2,$ we keep powers $\left(x/N \right)^{c-1}$ and $\left(x/N\right)^1.$ The order of the neglected terms depends on the asymptotic behavior of ${\cal U}_s\left(\ell\right).$ For ${\cal U}_s\left(\ell\right) \sim \ell^{-\hat{c}},$ $\hat{c} > c,$ the leading neglected term behaves as $\left(x/N\right)^{\min\left(2,c,\hat{c}-1\right)}.$ 

In expanding $\tilde{\cal U}\left(1 + x/N\right)$ we have to distinguish among $2 < c < 3$ and $1 < c < 2.$ (We will ignore the case $c = 2,$ where logarithmic corrections show up.)
\begin{itemize}
\item[$(i)$]{$2 < c < 3$}

We find
\begin{eqnarray}
\tilde{\cal U} \left(1 + \frac{x}{N}\right) & = & \tilde{\cal U}\left(1\right) + \sigma 
\Bigg[ \frac{\pi}{\Gamma \left(c\right) \sin \pi c} \left(\frac{x}{N}\right)^{c - 1} \nonumber \\
& & + \left(\left.\frac{d}{d\lambda}\right|_1 \tilde{\cal U}_s \left(\lambda\right) - \zeta \left(c-1\right)\right) \frac{x}{N} + \cdots  \Bigg] ~, 
\end{eqnarray}
where $\zeta\left(z\right)$ denotes the Zeta function. We now introduce a scaled `temperature like' variable $\bar\tau$ as
\begin{equation}
w = w^*\left(1 + \frac{\bar\tau}{a_\tau N}\right)~,
\end{equation}
where the microstructure dependent amplitude $a_\tau$ will be given below. Substituting Eqs.~(2.27), (2.28) into Eq.~(2.5) for $G\left(\lambda\right)$ and exploiting the definition (2.8) of $w^*$ we find
\begin{equation}
G^{-1}\left(\lambda\right) = \frac{\lambda}{w} - 1 - \tilde{U} \left(\lambda\right) = \frac{1}{w^* a_\tau N} \left[x - \bar\tau - a_1 N^{2-c} x^{c-1}\right] ~, 
\end{equation}
where
\begin{eqnarray}
a_\tau & = & \left[1 + w^* \sigma\left(\left.\zeta \left(c-1\right) - \frac{d}{d\lambda} \right|_1 \tilde{\cal U}_s \left(\lambda\right) \right) \right]^{-1}~, \\ 
a_1 & = &\frac{\pi}{\Gamma \left(c\right) \sin \pi c} \: w^* \sigma a_\tau ~.
\end{eqnarray}
\item[$(ii)$]$1 < c < 2$

The expansion of $\tilde{\cal U}\left(1 + x/N\right)$ yields
\begin{eqnarray}
\nonumber \tilde{\cal U}\left(1 + \frac{x}{N}\right) = \tilde{\cal U}\left(1\right) - \sigma \Bigg[ & & \frac{\Gamma \left(2-c\right)}{c-1} \left(\frac{x}{N}\right)^{c-1} \\
& & + \frac{x}{N} \left(\left.I - \frac{1}{\Gamma\left(c\right) \left(2-c\right)} - \frac{d}{d\lambda}\right|_1 \tilde{\cal U}_s \left(\lambda\right) \right) + \cdots \Bigg]~,
\end{eqnarray}
where $I$ stands for the integral
\begin{eqnarray*}
I = \frac{1}{\Gamma\left(c\right)} \int\limits^1_0 dy \: \frac{1}{y^2} \left[\left(-\ln\left(1-y\right)\right)^{c-1} - y^{c-1} \right]~.
\end{eqnarray*}
This time introducing $\bar\tau$ as
\begin{equation}
w = w^* \left(1 + \frac{\bar\tau}{a_\tau N^{c-1}}\right)
\end{equation}
we find for $G\left(\lambda\right)$
\begin{eqnarray*}
G^{-1}\left(\lambda\right) = \frac{N^{1-c}}{w^* a_\tau} \left[x^{c-1} - \bar\tau - a_1 N^{c-2} x\right]~,
\end{eqnarray*}
where
\begin{eqnarray}
a_\tau & = & \frac{c - 1}{w^* \sigma \Gamma \left(2 - c\right)} \\
a_1 & = & -a_\tau + a_\tau w^* \sigma \Bigg[\left. \frac{d}{d\lambda} \right|_1 \tilde{\cal U}_s \left(\lambda\right) + \frac{1}{\left(2 - c\right) \Gamma\left(c\right)} - I\Bigg]
\end{eqnarray}
Using the exponent $\phi,$ Eq.~(2.9), we can combine the results for $2 < c < 3$ and $1 < c < 2$ in the form
\begin{eqnarray}
\bar\tau & = & a_\tau \left(\frac{w}{w^*}-1 \right) N^{\phi}~, \\
G\left(\lambda\right) & = & w^* a_\tau N^\phi \bar{G}\left(x\right)~,\\
\bar{G}\left(x\right) & = & \left[x^\phi - \bar\tau - a_1 N^{-|c-2|} x^{c-\phi} \right]^{-1}~.
\end{eqnarray}
Turning now to ${\cal Z}_{bb},$ Eq.~(2.6), we note that with $\lambda = 1 + x/N$ we can replace the factor $\lambda^N$ by $e^x.$ Thus, to the order considered here we find
\begin{eqnarray}
{\cal Z}_{bb} \left(N\right) & = & w^*a_\tau N^{\phi-1} \bar{\cal Z}_{bb} \left(\bar\tau , N \right) ~, \\
\bar{\cal Z}_{bb} \left(\bar\tau, N\right) & = & \oint\frac{dx}{2 \pi i} \: e^x \bar{G}\left(x\right)~. 
\end{eqnarray}
The density of bound pairs takes the form
\begin{eqnarray}
\rho_{bb} & = & a_\tau N^{\phi-1} \bar\rho_{bb} \left(\bar\tau, N\right)~, \\
\bar\rho_{bb} \left(\bar\tau, N\right) & = & \frac{\partial}{\partial \bar\tau} \ln \bar{\cal Z}_{bb} \left(\bar\tau, N\right)~.
\end{eqnarray}
The melting curve is found as
\begin{eqnarray}
C_{bb} & = & a^2_\tau N^{2 \phi-1} \bar{C}_{bb}\left(\bar\tau, N\right) ~, \\
\bar{C}_{bb} \left(\bar\tau, N\right) & = & \frac{\partial}{\partial \bar\tau} \bar\rho_{bb} \left(\bar\tau, N\right)~.
\end{eqnarray}
The mean length of the first loop, Eqs.~(2.19), (2.20), involves a factor
\begin{eqnarray*}
\lambda \frac{d}{d \lambda} \tilde{\cal U}\left(\lambda\right) \Rightarrow N \frac{d}{dx} \left[\tilde{\cal U}\left(1 + \frac{x}{N}\right) - \tilde{\cal U}\left(1\right) \right]~.
\end{eqnarray*}
Calculating this factor for $c > 2$ or $c < 2$ from the expressions given above we find
\begin{equation}
\left<\ell_1\right>_{bb} = \frac{1}{a_\tau} \: \frac{1 - a_\tau}{1 - w^*} - \frac{c-1}{1-w^*} \: \frac{a_1}{a_\tau} \: N^{2-c} \: \frac{\bar{\cal Z}^{[\ell_1]}_{bb}}{\bar{\cal Z}_{bb}}~, \quad c > 2 ~,
\end{equation}
or
\begin{equation}
\left<\ell_1\right>_{bb} = \frac{c-1}{1-w^*} \: \frac{1}{a_\tau} N^{2-c}  \: \frac{\bar{Z}^{[\ell_1]}_{bb}}{\bar{\cal Z}_{bb}} \: -  \: \frac{1}{a_\tau}  \: \frac{a_1 + a_\tau}{1-w^*}, \: c < 2~,
\end{equation}
respectively. We introduced the function
\begin{equation}
\bar{\cal Z}^{[\ell_1]}_{bb} \left(\bar\tau; N\right) = \oint \frac{dx}{2 \pi i} \: e^x x^{c-2} \bar{G}\left(x\right)~,
\end{equation}
occuring in both expressions.

Turning now to $(bu)$-boundary conditions we have to specify the weight $p\left(m,N\right)$ of the open ends. We recall that $p\left(m,N\right)$ equals $1$ in the 
original formulation of the P.S.-model, but the network scenario suggests \cite{Z13} that open ends of length $1 \ll m \ll N$ should be weighted by a factor $\sim m^{-c_1}.$ Furthermore, as has been pointed out in Ref.~\cite{Z3}, also configurations $1 \ll N-m \ll N,$ where the closed part is small compared to the dangling ends, should be weighted by a factor $\left(N-m\right)^{-c_1}.$ As entropic weight for the end pieces we thus choose
\begin{equation}
p \left(m,N \right) = m^{-c_1} \left(N-m \right)^{-c_1}~.
\end{equation}

Some comment on this choice seems appropriate. It is found that with $(bu)$-boundary conditions in the critical region typically a sizeable part of the chain is open, so that using an asymptotic expression for $p \left(m,N\right)$ seems justified. However, the asymptotic power laws do not uniquely fix the form of $p \left(m, N \right).$ The form (2.48) is just the simplest ansatz. It has the virtue of introducing no additional parameters. We further should note that it has been argued \cite{Z3} that the power laws in $m$ or $(N-m)$ might involve different exponents. This `co-polymer network' scenario was motivated by an attempt to reproduce the measured effective exponents. Our analysis reveals that all effects discussed in this context can be traced back to the corrections to the finite size limit, embodied in the terms proportional to $N^{-|2 - c|},$ (see Sect.~\ref{kap4}). Thus there is no need of introducing more exponents, and no experimental evidence for a `co-polymer network' scenario is left \cite{Z14}.

Having specified $p\left(m,N\right)$ we can calculate ${\cal Z}_{bu}$ from Eq.~(2.21). We note that ${\cal Z}_{bb}\left(N-m\right)$ is given by Eqs.~(2.39), (2.40), with the only change that the factor $e^x$ in Eq.~(2.40) is to be replaced by $e^{x\left(1-\bar{m}\right)},$ where
\begin{equation}
\bar{m} = \frac{m}{N}~.
\end{equation}
Replacing the summation over $m$ by integration over $\bar{m},$ we immediately find
\begin{eqnarray}
{\cal Z}_{bu}\left(N\right) & = & w^* a_\tau N^{\phi-2c_1} \bar{\cal Z}_{bu}\left(\bar\tau, N\right)~, \\
\bar{\cal Z}_{bu} \left(\bar\tau, N\right) & = & \int\limits^1_0 d\bar{m} \:  \bar{m}^{-c1} \left(1-\bar{m}\right)^{-c_1} \oint \frac{dx}{2 \pi i} \: e^{x\left(1-\bar{m}\right)} \bar{G}\left(x\right)~.
\end{eqnarray}
The corresponding expression holds for $\bar{\cal Z}^{[\ell_1]}_{bu}.$ All the other results of the present subsection stay unchanged, with the index $(bb)$ replaced by $(bu).$
\end{itemize}

\subsection{Distribution functions}\label{kap2.3}

The results of the previous subsection immediately yield the distribution of the length $m$ of the open ends.
\begin{eqnarray}
{\cal P}^{[E]}_{bu} \left( m, N\right) = p \left(m, N \right) \frac{{\cal Z}_{bb} \left(N-m\right)}{{\cal Z}_{bu}\left(N\right)}~.
\end{eqnarray}
Eqs.~(2.39), (2.40), (2.49-2.51) result in
\begin{eqnarray}
{\cal P}^{[E]}_{bu} \left(m,N \right) & = & \frac{1}{N} \bar{\cal P}^{[E]}_{bu} \left(\bar{m}, \bar\tau, N \right)~, \\
\bar{\cal P}^{[E]}_{bu} \left(\bar{m}, \bar\tau, N \right) & = &  \frac{\bar{m}^{-c_1}\left(1-\bar{m}\right)^{-c_1}}{\bar{\cal Z}_{bu}\left(\bar\tau, N\right)} \oint\frac{dx}{2 \pi i} \: e^{x\left(1-\bar{m}\right)} \bar{G}\left(x\right)~.
\end{eqnarray}

Also the distribution of the loop length, averaged over all loops, is easily calculated. Up to normalization it for $(bb)$-boundary conditions is defined as 
\begin{eqnarray}
X_{bb} \left(\ell, N\right) = \sum\limits^{\infty}_{k = 0} \:\: & &\sum_{\{j_0, \cdots ,j_k\} \ge 1 \atop \{\ell_1, \cdots ,\ell_k\} \ge 1} \delta \Big(N - j_0 - \sum^{k}_{\mu = 1}\left(j_\mu + \ell_\mu\right)\Big) \nonumber \\
& & \sum^{k}_{\mu = 1} \delta_{\ell_\mu, \ell} \: w^{j_0} \prod\limits^{k}_{\mu = 1} \left({\cal U} \left(\ell_\mu \right) w^{j_n} \right)~.
\end{eqnarray}
The standard analysis yields in the critical region

\begin{eqnarray}
X_{bb} \left(\ell, N\right) & = & \left(w^*a_\tau\right)^2 N^{2 \phi -1} {\cal U} \left(\ell \right) \Theta \left(1 -\bar{\ell}\right) \bar{X}_{bb} \left(\bar\ell, \bar\tau, N\right)~, \\
\bar{X}_{bb} \left(\bar\ell, \bar\tau, N \right) & = & \oint \frac{dx}{2 \pi i} \: e^{\left(1 - \bar\ell\right) x} \bar{G}^2\left(x\right)~,
\end{eqnarray}
where
\begin{equation}
\bar\ell = \frac{\ell}{N}~,
\end{equation}
and $\Theta\left(z\right)$ is the step function. The normalization takes the form
\begin{equation}
\sum\limits_{\ell} X_{bb} \left(\ell, N \right) = w^* \left(1 - w^* \right) a^2_\tau N^{2 \phi - 1} \bar{X}_{bb}\left(0, \bar\tau, N \right)~.
\end{equation}
We thus find the loop length distribution
\begin{equation}
{\cal P}^{[LL]}_{bb}\left(\ell, N \right) = \frac{w^*}{1 - w^*} \: {\cal U} \left(\ell \right) \Theta \left(1 - \bar\ell \right) \frac{\bar{X}_{bb}\left(\bar\ell, \bar\tau,N\right)}{\bar{X}_{bb}\left(0, \bar\tau,N\right)}~.
\end{equation}
For $(bu)$-boundary conditions $\bar{X}_{bb}$ is to be replaced by
\begin{eqnarray}
\bar{X}_{bu} \left(\bar\ell, \bar\tau, N \right) = \int\limits^{1-\bar\ell}_{0} d \bar{m} \: \bar{m}^{-c_1} \left(1 - \bar{m}\right)^{-c_1} \oint \frac{dx}{2 \pi i} \exp \left(\left(1 - \bar{m} - \bar\ell\right) x\right) \bar{G}^2\left(x\right)~. 
\end{eqnarray}

Analysis of the distribution of the number of bound pairs is somewhat more involved. For $(bb)$-boundary conditions it is defined as
\begin{equation}
{\cal P}^{\left[ BP \right]}_{bb} \left(n, N \right) = \frac{{\cal Y}_{bb}\left(n,N\right)}{{\cal Z}_{bb}\left(N\right)}~, 
\end{equation}
\begin{eqnarray}
{\cal Y}_{bb} \left(n,N \right) = \sum\limits^\infty_{k=0} & & \sum\limits_{\{j_0, \cdots ,j_k\} \ge 1 \atop \{\ell_1, \cdots ,\ell_k\} \ge 1} \delta \Big(N - j_0 - \sum^k_{\mu = 1} \left(j_\mu + \ell_\mu \right)\Big) \nonumber \\
& & \delta  \Big(n - \sum^k_{\mu = 0} j_\mu \Big) w^{j_0} \prod\limits^k_{\mu=1} \left({\cal U} \left(\ell_\mu\right) w^{j_\mu}\right)~.
\end{eqnarray}
${\cal Y}_{bb}$ again can be evaluated by introducing a generating function, resulting in
\begin{equation}
{\cal Y}_{bb}\left(n, N\right) = w^n \oint \frac{d\lambda}{2 \pi i\lambda} \lambda^{N-n} \left[1 + \tilde{\cal U} \left(\lambda\right)\right]^{n-1}~.
\end{equation}
The integral encircles the cut in the $\lambda$-plane. We are interested in the range $n \gg 1,$ $N-n \gg 1,$ since only there we can expect to find results independent of microscopic details of the model. In this region the integral is dominated by the branch point $\lambda^* = 1.$ We thus again introduce variables $\bar\tau$ and $x$ via Eqs.~(2.36), (2.26), and we use Eq.~(2.8) to write
\begin{eqnarray*}
1 + \tilde{\cal U}\left(\lambda\right) = \frac{1}{w^*} \left[ 1 - w^* \left(\tilde{\cal U} \left(1\right) - \tilde{\cal U} \left( 1 + \frac{x}{N} \right)\right) \right] ~.
\end{eqnarray*}
We furthermore introduce the scaled variable
\begin{equation}
\bar{n} = \frac{n}{a_\tau} N^{-\phi} ~.
\end{equation}
Eq.~(2.64) takes the form
\begin{eqnarray}
{\cal Y}_{bb} \left(n,N\right) = \frac{w^*}{N} \left(1 + \frac{\bar\tau}{a_\tau N^\phi}\right)^{a_\tau N^{\phi} \bar{n}} \oint \frac{dx}{2 \pi i} \left(1 + \frac{x}{N}\right)^{N\left(1-a_\tau \bar{n}N^{\phi-1}\right)-1} \nonumber \\ 
\cdot \left[1 - w^* \left(\tilde{\cal U} \left(1\right) - \tilde{\cal U} \left(1 + \frac{x}{N}\right) \right) \right]^{a_\tau \bar{n} N^{\phi}-1}
\end{eqnarray}
Using the expansion of $\tilde{\cal U} \left(1 + \frac{x}{N}\right),$ and invoking $N \gg 1$ we find
\begin{eqnarray}
{\cal Y}_{bb}\left(n, N\right) & = & \frac{w^*}{N} e^{\bar\tau \bar{n}} \bar{\cal Y}_{bb}\left(\bar{n},N \right)~, \\ 
\bar{\cal Y}_{bb}\left(\bar{n}, N\right) & = & \oint \frac{dx}{2 \pi i}\exp \left[ x - x^\phi \bar{n} + a_1 \bar{n} N^{-|c-2|} x^{c-\phi} \right]~.
\end{eqnarray}
Taking into account the normalization ${\cal Z}_{bb}\left(N\right),$ Eq.~(2.39), we find as final result
\begin{eqnarray}
{\cal P}^{\left[ BP \right]}_{bb} \left(n, N \right) & = & \frac{1}{a_\tau N^\phi} \bar{\cal P}^{\left[ BP \right]}_{bb} \left(\bar{n}, \bar\tau, N \right)~, \\
\bar{\cal P}^{\left[ BP \right]}_{bb} \left(\bar{n}, \bar\tau, N \right) & = & e^{\bar\tau \bar{n}} \frac{\bar{\cal Y}_{bb}\left(\bar{n},N\right)}{\bar{\cal Z}_{bb}\left(\bar\tau, N \right)}~.
\end{eqnarray}
For $(bu)$-boundary conditions Eqs.~(2.69), (2.70) hold with the index $(bb)$ replaced by $(bu),$ where $\bar{\cal Y}_{bu} \left(\bar{n},N\right)$ is defined as
\begin{eqnarray}
\bar{\cal Y}_{bu}\left(\bar{n},N\right) =  \int\limits^{1-a_\tau \bar{n} N^{\phi-1}}_{0} d\bar{m} \: \bar{m}^{-c_1} \left(1-\bar{m}\right)^{-c_1} & & \oint \frac{dx}{2 \pi i} \exp \Big[\left(1-\bar{m}\right) x \nonumber \\
& & - \bar{n} x^{\phi} + a_1 \bar{n} N^{-|c-2|} x^{c-\phi} \Big]~.
\end{eqnarray}

\newpage

\section{Finite size scaling limit}\label{kap3}

Taking the limit $N \to \infty$ with the scaling variables $\bar\tau, \bar{n}, \bar\ell, \bar{m}$ held fixed just amounts to dropping the terms $a_1 N^{-|c-2|} x^{c-\phi}$ in all expressions of the previous section. We consider the cases $c > 2$ or $c < 2$ separately.

\subsection{$c > 2:$ $\phi = 1$}\label{kap3.1}
For $N \to \infty$ the contribution of the cut vanishes in the expressions for the partition function, Eqs.~(2.40), (2.51). For all $\bar\tau$ only a simple pole survives. We thus for $(bb)$ boundary conditions find the simple result
\begin{equation}
\hat{\cal Z}_{bb} \left(\bar\tau\right) = \lim_{N \to \infty} \bar{\cal Z}_{bb} \left(\bar\tau,N\right) = e^{\bar\tau}~.
\end{equation}
This yields a constant density of bound pairs,
\begin{equation}
\rho_{bb} = a_\tau~,
\end{equation}
and vanishing fluctuations,
\begin{equation}
C_{bb} = 0~.
\end{equation}
Indeed, the distribution of the number of bound pairs degenerates to a $\delta$-function.
\begin{equation}
{\cal P}^{[BP]}_{bb} \left(n,N \right) = \frac{1}{N} \delta \left(a_\tau - \frac{n}{N} \right)
\end{equation}

To derive Eq.~(3.4) some closer inspection of Eq.~(2.68) for $\bar{\cal Y}_{bb} \left(\bar{n}, N \right)$ is necessary.
\begin{eqnarray*}
\bar{\cal Y}_{bb} \left(\bar{n}, N \right) & = & \oint \frac{dx}{2\pi i} \exp \left[ x \left(1 - \bar{n} \right) + a_1 \bar{n} N^{-(c-2)} x^{c-1}\right]~, \\
\bar{n} & = & \frac{1}{a_\tau} \: \frac{n}{N}~.
\end{eqnarray*}
We first note that $a_\tau,$ Eq.~(2.30), obeys the inequality
\begin{equation}
0 < a_\tau < 1~. 
\end{equation}
This is immediately obvious from Eq.~(3.2) and formally derives from the properties of the generating function $\tilde{\cal U} \left(\lambda\right).$ Thus also $a_1,$ Eq.~(2.31), is positive. A priory the integral for $\bar{\cal Y}_{bb}$ extends over the edge of the cut $x < 0,$ and for $n/N > a_\tau$ the term proportionally to $a_1$ is necessary to guarantee convergence. Since $c > 2$ we, however, can transform the integration contour to the imaginary axis, where the limit $N \to \infty$ can be taken. The result
\begin{eqnarray*}
\hat{\cal Y}_{bb} \left(\bar{n}\right) = \lim_{N\to\infty} \bar{\cal Y}_{bb} \left(\bar{n}, N\right) = \delta \left( 1- \bar{n}\right)
\end{eqnarray*}
follows.

For the distribution of loop lengths, Eq.~(2.60), the finite size scaling limit is also easily evaluated.
\begin{equation}
{\cal P}^{[LL]}_{bb} \left(n,N\right) 
\mathop{\,\hbox to 2.5em{$-$\hskip-1ex\leaders\hbox to1ex{\hss$-$\hss}\hfill
        \hskip-1ex$\longrightarrow$}\,}\limits_{N\to\infty}
 \frac{w^*}{1-w^*} \:{\cal U}\left(\ell\right) \Theta \left(1-\bar\ell\right) \left(1-\bar\ell\right) e^{-\bar\ell \bar\tau}
\end{equation}
Eqs.~(2.17), (3.2) yield a constant loop density, $\rho^{[L]}_{bb} = \left( 1-w^*\right) a_\tau.$ Also the mean length of the first loop, Eq.~(2.45), tends to a constant:
\begin{equation}
\left<\ell_1\right>_{bb} = \frac{1}{a_\tau} \: \frac{1 - a_\tau}{1-w^*}~.
\end{equation}
It in fact becomes identical to the mean length averaged over all loops, Eq.~(2.18).

In summary, we have found that for $c > 2$ in the finite size scaling limit the internal structure of the chain bound together at both ends is independent of $\bar\tau$ in the whole critical region $- \infty < \bar\tau < \infty.$ A priory this is somewhat surprising. The explanation is provided by Eq.~(3.6), which for $\bar\ell > 0,$ $\bar\tau = \mbox{const},$ $N \gg 1$ properly should be written as
\begin{eqnarray*}
{\cal P}^{[LL]}_{bb} \sim N^{-c} \bar\ell^{-c} \left(1-\bar\ell\right) e^{-\bar\ell \bar\tau}~.
\end{eqnarray*}
The total length contained in loops $\bar\ell > \eta$ for any $\eta > 0$ becomes non-negligible only for
\begin{eqnarray*}
\bar\tau \sim - \ln N
\mathop{\,\hbox to 2.5em{$-$\hskip-1ex\leaders\hbox to1ex{\hss$-$\hss}\hfill
        \hskip-1ex$\longrightarrow$}\,}\limits_{N\to\infty}
-\infty~.
\end{eqnarray*}
Thus denaturation occurs outside the finite size scaling limit as defined above.

If the chain is allowed to open from one end we find a completely different behavior. Eq.~(2.51) yields
\begin{eqnarray}
\hat{\cal Z}_{bu}\left(\bar\tau\right) & = & \int\limits^{1}_{0} d \bar{m} \left(\bar{m}\left( 1 -\bar{m}\right)\right)^{-c_1} e^{\left(1-\bar{m}\right)\bar\tau} \nonumber \\
& = & \sqrt{\pi} \Gamma \left(1 - c_1\right) \bar\tau^{c_1 - 1/2} e^{\bar\tau/2} I_{\frac{1}{2}-c_1} \left(\bar\tau/2 \right)~,
\end{eqnarray}
where $I_\nu\left(z\right)$ is the modified Bessel function. The density of bound pairs and the associated fluctuations take the form
\begin{eqnarray*}
\rho_{bu} & = & a_\tau \hat\rho_{bu} \left(\bar\tau\right) \\
C_{bu} & = & a^2_\tau N \hat{C}_{bu}\left(\bar\tau\right)~,
\end{eqnarray*}
with scaling functions
\begin{eqnarray}
\hat\rho_{bu} \left(\bar\tau\right) & = & \frac{1}{2} + \frac{1}{2} \, \frac{I_{\frac{3}{2}-c_1} \left(\bar\tau/2\right)}{I_{\frac{1}{2}-c_1}\left(\bar\tau /2 \right)}~, \\
\hat{C}_{bu} \left(\bar\tau\right) & = & \frac{1}{4} - \frac{1}{4} 
\left(\frac{I_{\frac{3}{2}-c_1} \left(\bar\tau /2 \right)}{I_{\frac{1}{2}-c_1}\left(\bar\tau /2 \right)}\right)^2
- \frac{1-c_1}{\bar\tau} \, \frac{I_{\frac{3}{2}-c_1} \left(\bar\tau/2\right)}{I_{\frac{1}{2}-c_1}\left(\bar\tau /2 \right)}~.
\end{eqnarray}
We note that $\hat{C}_{bu}\left(\bar\tau\right)$ is symmetric in $\bar\tau:$
\begin{eqnarray*}
\hat{C}_{bu}\left(-\bar\tau\right) = \hat{C}_{bu}\left(\bar\tau\right)~.
\end{eqnarray*}

With the choice $c_1 = 0.11$ suggested by the network scenario the scaling functions are plotted in Figs.~1a or 2a, respectively. It must be stressed that in view of the results for $(bb)$-boundary conditions the observed $\bar\tau$-dependence is due only to the progressive prolongation of the dangling ends. It is easily checked that $\hat\rho_{bu}\left(\bar\tau\right)$ is related to the average length $<m/N>$ of the open ends as
\begin{equation}
\hat\rho_{bu}\left(\bar\tau\right) = 1 - \left< \frac{m}{N}\right>~.
\end{equation}

Also the distribution of the number of bound pairs is easily evaluated. Following the steps explained above in the context of the evaluation of $\bar{\cal Y}_{bb}$ we find
\begin{eqnarray}
\hat{\cal P}^{[BP]}_{bu} \left(\bar{n}, \bar\tau \right) & = & \lim_{N \to \infty} \bar{\cal P}^{[BP]}_{bu} \left(\bar{n}, \bar\tau, N \right) \nonumber \\ 
& = & \left( \left(1-\bar{n}\right) \bar{n}\right)^{-c_1} \Theta \left(1-\bar{n}\right) \frac{e^{\bar\tau \bar{n}}}{\hat{\cal Z}_{bu}\left(\bar\tau\right)}~.
\end{eqnarray}
Thus this distribution at the critical point $\bar\tau = 0,$ (Fig.~3a), just reflects our choice of $p\left(m,N \right).$ $\hat{\cal P}^{[BP]}_{bu}$ is directly related to the scaling function of the length distribution of the dangling ends:
\begin{equation}
\hat{\cal P}_{bu} \left(\bar{n}, \bar\tau\right) = \hat{\cal P}^{[E]}_{bu} \left(1 - \bar{n}, \bar\tau \right)~.
\end{equation}
Thus for $\bar\tau = 0$ the length of the open ends fluctuates strongly. For $c_1 = 0$ its distribution would be completely flat. The choice $c_1 > 0$ induces a weak preference of essentially open or essentially closed configurations.

To summarize, we have found a first important result. For $(bu)$-boundary conditions and $c > 2$ denaturation in the finite size scaling limit is completely due to the opening of the end. The closed part of the chain is just a passive spectator, undergoing no relevant change in its loop structure. The numerical value of $c > 2$ therefore is irrelevant.

\subsection{$1 < c < 2:$ $\phi = c - 1$}\label{kap3.2}
In the region of the second order transition the contribution of the cut survives the finite size scaling limit, and the results become less trivial. Eq.~(2.40) yields
\begin{eqnarray}
\hat{\cal Z}_{bb}\left(\bar\tau \right) & =  & \oint \frac{dx}{2 \pi i} \: \frac{e^x}{x^{c-1}- \bar\tau} \nonumber \\
& = & - \: \frac{\sin \pi c}{\pi} \int\limits^{\infty}_{0}dx \frac{x^{c-1}e^{-x}}{\left(x^{c-1} \cos \pi c + \bar\tau\right)^2 + \left(x^{c-1}  \sin \pi c\right)^2} \nonumber \\
& & + \: \frac{\Theta \left(\bar\tau\right)}{c -1} \bar\tau^{\frac{2-c}{c-1}} \exp \left(\bar\tau^{\frac{1}{c-1}} \right)~.
\end{eqnarray}
For $(bu)$-boundary conditions we find from Eq.~(2.51)
\begin{eqnarray}
\hat{\cal Z}_{bu}\left( \bar\tau\right) & = & - \: \frac{\sin \pi c}{\sqrt{\pi}} \Gamma \left(1 - c_1 \right) \int\limits^{\infty}_{0} dx \frac{x^{c + c_1 -3/2}e^{-x/2} I_{1/2 - c_1} \left(x/2\right)}{\left(x^{c-1} \cos \pi x + \bar\tau\right)^2 + \left(x^{c-1} \sin \pi c \right)^2} \nonumber \\
& & + \: \frac{\Theta \left(\bar\tau\right)}{c-1} \sqrt{\pi} \Gamma \left(1 - c_1 \right) \bar\tau^{\frac{1}{c-1}\left(3/2 - c + c_1\right)} \exp\left(\frac{\bar\tau^{\frac{1}{c-1}}}{2} \right) I_{\frac{1}{2}-c_1} \left(\frac{1}{2}\bar\tau^{\frac{1}{c-1}}\right)~.
\end{eqnarray}
Clearly the resulting scaling functions $\hat\rho \left(\bar\tau\right)$ and $\hat{C} \left(\bar\tau\right)$ have to be evaluated numerically. (As a technical remark we note that for $\bar\tau \approx 0$ it is preferable to include in the integration contour for $x$ a small circle centered at $x = 0,$ so as to avoid the numerical complications of the pole merging with the branch point.) The results for $c = 1.75,$ $c_1 = 0$ are shown in Figs.~1b or 2b, respectively. We see some quantitative effect of the boundary conditions but qualitatively the behavior is not changed.

However, an analysis of the bond number distribution again shows a strong sensitivity to the boundary conditions. On the technical side we note that the integral (2.68) for $\bar{\cal Y}_{bb}\left(\bar{n}, N \right)$, $c < 2,$ best is evaluated along a path of constant phase in the $x-$plane. In polar coordinates $(r, \psi)$ this path for $N \to \infty$ is parameterized as
\begin{equation}
r \left(\psi, \bar{n}\right) = \left[\bar{n} \frac{\sin \left(c-1\right)\psi}{\sin \psi} \right]^{\frac{1}{2-c}}~,
\end{equation}
and $\hat{\cal Y}_{bb}\left(\bar{n}\right)$ takes the form
\begin{eqnarray}
\hat{\cal Y}_{bb}\left(\bar{n}\right) = \frac{\left(c-1\right)\bar{n}}{\left(2-c\right)\pi} & & \int\limits^{\pi}_{0} d \psi r^{c-1}\left(\psi, \bar{n}\right) \frac{\sin\left(2-c \right) \psi}{\sin \psi} \nonumber \\
& & \cdot \exp \left[ - \bar{n} r^{c-1}\left(\psi, \bar{n}\right) \frac{\sin\left(2-c \right) \psi}{\sin \psi} \right]~.
\end{eqnarray}
A similar expression is found for $\hat{\cal Y}_{bu}\left(\bar{n}\right).$ For $\bar\tau = 0,$ $c = 1.75,$ $c_1 = 0$ the resulting bond-number distributions are shown in Fig.~3b. The qualitative difference among $\hat{\cal P}^{[BP]}_{bb}$ and $\hat{\cal P}^{[BP]}_{bu}$ reflects the fluctuations in the length of the dangling ends, which again are very strong. Evaluating Eq.~(2.54), for $c < 2,$ $c_1 = 0,$ $\bar\tau = 0,$ $N \to \infty,$ we find the simple result
\begin{equation}
\hat{\cal P}^{[E]}_{bu} \left(\bar{m}, 0\right) = \left(c-1\right) \left(1 - m \right)^{c-2}~.
\end{equation}
Again this distribution is quite flat, and the most probable configuration is an essentially open chain. Thus also for $c < 2$ the denaturation transition strongly is driven by the opening of the chain ends.

We finally recall that the density of loops just follows the density of bound pairs and thus strongly decreases with decreasing $\bar\tau,$ whereas the mean length averaged over all loops, Eq.~(2.18), increases. $\left<\ell\right>$ is also sensitive to $N.$ Evaluating $\rho_{bb}$ for $\bar\tau = 0,$ $N \to \infty,$ we find
\begin{equation}
\rho_{bb}\left(w^*, N\right) = a_\tau N^{c-2} \frac{\Gamma\left(c-1\right)}{\Gamma\left(2c-2\right)}~,
\end{equation} 
resulting in
\begin{equation}
\left<\ell\right>_{bb} = \frac{N^{2-c}}{a_\tau \left(1-w^*\right)} \: \frac{\Gamma\left(2c-2\right)}{\Gamma\left(c-1\right)}~, \: \bar\tau = 0~.
\end{equation}
This is to be compared to the mean length of the first loop, which from Eq.~(2.46) is found as
\begin{equation}
\left<\ell_1\right>_{bb} = \frac{N^{2-c}}{a_\tau\left(1-w^*\right)} \: \Gamma\left(c\right)~, \: \bar\tau = 0~.
\end{equation}
As expected, $\left<\ell_1\right>$ asymptotically exceeds $\left<\ell\right>.$ Specifically for $c=1.75$ we find
\begin{eqnarray*}
\lim_{N\to\infty \atop \bar\tau = 0} \frac{\left<\ell_1\right>_{bb}}{\left<\ell\right>_{bb}} \approx 1.27~.
\end{eqnarray*}

\newpage

\section{A fit to the results of simulations}\label{kap4}

The simple model introduced and simulated in Ref.~\cite{Z1} briefly has been described in the introduction. We recall that it uses $(bu)$-boundary conditions for the two strands of the chain and involves a single parameter $e^\epsilon$ weighting the overlap of complementary base pairs. Except for allowing for such overlaps the excluded volume is fully incorporated.

In Ref.~\cite{Z1} data for the density of bound pairs $\rho_{bu}$ and the specific heat $\epsilon^2 C_{bu}$ are shown as function of $\epsilon$ for single strand lengths up to $N = 3000.$ In addition, for three values of $\epsilon$ close to the estimated critical value $\epsilon^* = 1.3413$ the distribution functions ${\cal P}^{[BP]}_{bu}$ of the number of bound pairs are given. Analysis of these distributions, in particular, led the authors to conclude that the transition is of first order, but sizeable corrections to finite size scaling exist.

For the same model data on the loop length distribution ${\cal P}^{[LL]}_{bu}$ for chains up to length $N = 1280$ at $\epsilon = 1.3413$ have been presented in Refs.~\cite{Z3,Z4}. The authors extract a value of $c = 2.14 \pm 0.04.$ Ref.~\cite{Z3} also provides some information on the distribution of the length of the dangling ends, ${\cal P}^{[E]}_{bu},$ and on the chain length dependence of the partition function ${\cal Z}_{bu}.$ Other data presented in Refs.~\cite{Z3,Z1} concern spatial properties like the distance distribution within a loop, or refer to the phase diagram in the limit where the chain fills a nonvanishing fraction of the volume. Such aspects are outside the frame of the present work.

Inspection of the simulation results immediately shows that even for the longest chains the data cannot be reproduced by the finite size scaling limit of the P.S.-model. In particular, the measured distribution ${\cal P}^{[BP]}_{bu}\left(n/N\right),$ $\epsilon = 1.3413,$ only vaguely resembles the theoretically predicted ${\hat{\cal P}}^{[BP]}_{bu},$ $\bar\tau \approx 0,$ $c >2,$ (Fig.~3a), and differs completely from $\hat{\cal P}^{[BP]}_{bu},$ $c < 2,$ (Fig.~3b). Whether including the corrections $\sim a_1 N^{-|c-2|}$ allows for fitting the data is the topic considered here. For given exponents $\left(c, c_1\right)$ this introduces $a_1$ as a second fit parameter besides $a_\tau.$ A third and quite important fit parameter is the critical value $\epsilon^*.$ Clearly $e^\epsilon$ is proportional to the parameter $w$ introduced in the P.S.-model. In the critical region we write
\begin{eqnarray*}
\frac{w}{w^*} = e^{\epsilon - \epsilon^*} \approx 1 + \epsilon - \epsilon^*~.
\end{eqnarray*}
so that Eq.~(2.36) for the scaling variable $\bar\tau$ yields
\begin{equation}
\bar\tau = a_\tau \left(\epsilon - \epsilon^* \right) N^\phi~,
\end{equation}
Since with the present data $N^\phi$ takes values up to $N^1 = 3000,$ even changing $\epsilon^*$ by $0.0001$ will have some effect. Note that in ${\cal P}^{[BP]}_{bu},$ $\bar\tau$ occurs in the argument of an exponential function, (cf.~Eq.~(2.70)).

In fitting we proceed as follows. For given $\left(c, c_1 \right)$ we determine $a_\tau, a_1, \epsilon^*$ by fitting simultaneously for $N = 500,$ $3000$ the results for the average bond number $\rho_{bu}$ as function of $\epsilon,$ given in Fig.~7a of Ref.~\cite{Z1}. We choose $\rho_{bu},$ since it shows the smallest statistical error. Analysis of the loop length distribution involves the prefactor $w^* \sigma/ \left(1 - w^*\right),$ (cf.~Eq.(2.60)), which is determined by fitting the results of Fig.~1, Ref.~\cite{Z4}. With these parameters known, the more directly interpretable parameters $w^*,$ $\sigma,$ $\tilde{\cal U} \left(1 \right),$ $\tilde{\cal U}^{\prime} \left(1 \right) = d/d\lambda \: \tilde{\cal U} \left(\lambda\right)\mid_1$ can be calculated. Concerning the numerical evaluation of our expressions the essential technical points have been mentioned in the previous section.

We now present our results for the choice $\left(c, c_1\right) = \left(2.05, 0.13 \right).$ These values are deep in the range of exponents which allow for a reasonable fit, as discussed in the next section. The fit uses parameters $\epsilon^* = 1.34110,$ $a_\tau = 0.2775,$ $a_1 = 0.6746,$ $w^* \sigma/\left(1-w^* \right) = 1.0,$ leading to $w^* \approx 0.87,$ $\sigma \approx 0.14,$ $\tilde{\cal U}_s\left(1 \right) \approx -0.60,$ $\tilde{\cal U}^{\prime}_s\left(1 \right) \approx -0.46.$ We note that $e^{\epsilon^*}/w^*$ plays the role of an effective coordination number of the lattice walk. The value $e^{\epsilon^*}/w^* \approx 4.4$ resulting from the above parameters seems quite reasonable for the cubic lattice used in the simulations. In fact it is quite close to the effective coordination number $\approx 4.68$ measured for a simple self-avoiding walk on that lattice.

Fig.~4 shows the average bond density and the melting curves as function of $\left(\epsilon - \epsilon^* \right)N$ for $N = 3000,$ $1000,$ $500.$ We omitted data for other chain lengths, so as not to overload the plots. For all $N$ the fits are of the same quality as those shown. Note that in panel $b)$ we have plotted $C_{bu}/N,$ since this is the quantity approaching a finite size scaling limit. We clearly can state excellent agreement among theory and data. Only for $N = 3000$ we observe a small shift of the maximum of the calculated melting curve relative to the data. However, this deviation is within the error bars, (cf. Fig.~8 of Ref.~\cite{Z1}), which we have suppressed here. The scaling limits, $N \to \infty,$ are included in Fig.~4 as broken lines. Obviously our fit implies that we are far from the scaling limit. With corrections decaying only as $N ^{-\left(c-2\right)} = N^{-0.05},$ this is no surprise.

We now turn to the distribution of the number of bound pairs, shown in Figs.~5, 6 of Ref.~\cite{Z1} for values $\epsilon = 1.3413,$ $\epsilon = 0.999 \cdot 1.3413 = 1.33996,$ $\epsilon = 1.001 \cdot 1.3413 = 1.34264.$ We first note that this distribution by construction of the simulation model factorizes according to 
\begin{equation}
{\cal P}^{[BP]}_{bu}\left(n, \epsilon, N \right) = \frac{{\cal Y}\left(n, N \right)}{{\cal Z}\left(\epsilon, N \right)} \: e^{\epsilon n}~,
\end{equation}
a property also valid in the P.S.-model, (cf.~Eq.~(2.70)). Thus
\begin{equation}
\frac{{\cal P}^{[BP]}_{bu}\left(n, \epsilon_1,N \right)}{{\cal P}^{[BP]}_{bu}\left(n, \epsilon_2,N \right)} \sim e^{\left(\epsilon_1 - \epsilon_2 \right)n}~.
\end{equation}
We have checked that the data fulfill this relation within about $3\%$ deviation. Thus measurements with different $\epsilon$ essentially carry the same information. Still, changing $\epsilon$ changes the weight associated with different regions of $n/N.$ We therefore in Fig.~5 show fits to data for the largest and the smallest value of $\epsilon.$ We note that we did not show the results of the theory for $n < 30.$ From the analysis of the finite size scaling limit we know that this region is dominated by essentially open chains with closed parts of length $\alt 100.$ For such short parts we expect the subleading corrections neglected here to become relevant. With this qualification we can state full agreement among theory and data. This strongly suggests that the P.S.-model indeed captures the essential physics of the problem.

Fig.~6 again illustrates that the present interpretation of the data implies that we are far from the finite size scaling limit. It compares ${\cal P}^{[BP]}_{bu}$ at the critical point $\bar\tau = 0$ as calculated for $N = 3000,$ (full line), to the finite size scaling limit, (long dashes). The result for $N = 3000$ decreases monotonically, showing no precursor of the singularity developing for $N \to \infty$ at $n/N = a_\tau.$ In the figure we also included the result for $\epsilon = 1.3413,$ $N = 3000,$ $\left(\bar\tau \approx 0.1665 \right),$ (short dashes). It shows a shallow maximum near $n/N = 0.4,$ which clearly is not related to the asymptotic singularity, but rather is due to the exponential prefactor $\exp \left( \bar{n} \bar\tau\right).$ This maximum has also been observed in the simulations, and the present interpretation contradicts that given in Ref.~\cite{Z1}.

We next consider the distribution of loop lengths. Data for $\epsilon = 1.3413$ and different chain lengths are presented in Ref.~\cite{Z3}, Fig.~7, and Ref.~\cite{Z4}, Fig.~1, as essentially continuous curves, from which we have drawn some points. We note that for the larger chain lengths the data curves for loop lengths $\ell \agt 100$ are rather noisy. We therefore for $N = 1280$ represent the data by a set of vertical bars, which give an impression of the range in which the data fluctuate. Fig.~7 shows our fit for chain lengths $N = 160,$ $320,$ $1280,$ using only the long range part $\sigma \ell^{-c}$ for the factor ${\cal U}\left(\ell \right)$ in Eq.~(2.60). Even taking into account that this plot is doubly logarithmic, we find the fit quite remarkable. Note that we have chosen the exponent $c = 2.05.$ Still the theory in some intermediate range $10 \alt \ell \alt 100$
yields an effective exponent $c \approx 2.14,$ as quoted in Ref.~\cite{Z4}. Obviously the variation of the factor $\bar{X}_{bu}\left(\bar{\ell}\right)$ in the expression for the loop length distribution is quite essential. Evaluating this factor for $\bar\tau = 0$ in the finite size scaling limit one finds that the asymptotic exponent $c$ within an error of about $.02$ can be extracted only from a region $1 \ll \ell \alt 10^{-2}N.$ Clearly, even for $N = 1280$ such a region does not exist. We also note that the present fit yields values $|\tilde{\cal U}_s \left(1 \right)| \approx 0.60,$ $|\tilde{\cal U}^{\prime}_s \left(1 \right)| \approx 0.46,$ that are fairly small compared to their long range counterparts $\sum^\infty_{\ell = 1} \ell^{-c} = \zeta \left(2.05\right) \approx 1.60,$ $\sum^\infty_{\ell = 1} \ell^{1-c} = \zeta \left(1.05\right) \approx 20.6.$ This is consistent with the observation that the fit works down to very small values of $\ell.$

For the distribution ${\cal P}^{[E]}_{bu}\left(m, N \right)$ of the length of the dangling ends data have been presented only for the two-dimensional version of the simulation model. (See Fig.~4 of Ref.\cite{Z3}.) Two distinct power laws have been extracted:
\begin{equation}
{\cal P}^{[E]}_{bu}\left(m, N \right) \sim 
\left\{ 
\begin{array}{r@{\quad=\quad}l}
m^{-{c^\prime}_1}, {c^\prime}_1 & 0.23 \pm 0.01; \: m \ll N \\
\left(N - m \right)^{-c_2}, c_2 & 0.35 \pm 0.01; \: N-m \ll N ~. 
\end{array} 
\right. 
\end{equation}
Both exponents differ from $c_1 = 9/32 \approx 0.28,$ predicted by the network scenario for $d = 2.$ For $d=3$ Ref.~\cite{Z3} quotes values ${c^\prime}_1 = 0.14 \pm 0.01,$ $c_2 = 0.16 \pm 0.01.$ As mentioned in Sect.~\ref{kap2.2} the occurrence of two different power laws led the authors to suggest a `co-polymer network' scenario, assuming that critical exponents for the bound part of the chain differ from those of the dangling ends.

Evaluating our expression for ${\cal P}^{[E]}_{bu}$ for finite $N$ we find that effective exponents defined as in Eq.~(4.4) always obey the relation ${c^\prime}_1 < c_1 < c_2.$ Indeed, using exponents $c = 77/32,$ $c_1 = 9/32,$ predicted by the simple network scenario in two dimensions, we easily can reproduce the measured effective exponents for the range of chain lengths considered in Ref.~\cite{Z3}. Turning to three dimensions we for the choice of parameters employed in this section in Fig.~8 show doubly logarithmic plots of $\bar{\cal P}^{[E]}_{bu}$ as function of $m/N$ or $1 - m/N,$ respectively. Pressed to extract effective exponents we would quote ${c^\prime}_1 \approx 0.12,$ $c_2 \approx 0.17,$ not far from the simulation results quoted above. In particular, $c_2$ definitely is larger than the value $c_1 = 0.13$ used and is in the range quoted from the simulations. ${c^\prime}_1$ is a little smaller than expected, but we doubt that such a small deviation is significant.

To support the co-polymer scenario, in Ref.~\cite{Z3} also the partition function itself has been measured. At the critical point ${\cal Z}\left( N\right)$ is expected to behave as
\begin{equation}
{\cal Z}\left(N \right) \sim N^{\gamma^*-1} e^{\mu N}~, 
\end{equation}
where the simple network scenario predicts $\gamma^* \approx 2.06$ in three dimensions, whereas with exponents ${c^\prime}_1,$ $c_2$ extracted from ${\cal P}^{[E]}_{bu}$ the co-polymer scenario leads to $\gamma^* \approx 2.00.$ The authors argue that their data favor the latter value. (See Fig.~8 of Ref.~\cite{Z3}.) Within the philosophy of the network approach we would write the partition function at $\bar\tau = 0$ as
\begin{equation}
{\cal Z}\left(N\right) \sim  N^{\gamma^*-1} e^{\mu N} \bar{\cal Z}_{bu}\left(0, N\right)~,
\end{equation}
an expression which for $N \to \infty$ reduces to Eq.~(4.5). Evaluating the correction factor $\bar{\cal Z}_{bu}\left(0, N\right)$ with the exponents and parameter values used in this section we find that in the range $100 \alt N \alt 1000$ it decreases the effectively measured exponent by about $0.06,$ thus bridging the gap between $\gamma^* = 2.06$ and the value $\gamma^* = 2.00$ advocated in Ref.~\cite{Z3}. In summary, we in this section have shown that the P.S.-model, slightly generalized by a simple end-weighting factor, allows for a consistent and excellent fit of the available Monte Carlo data. Keeping the leading corrections to the finite size scaling limit is essential. For the chain length measured these corrections explain all the observed effective exponents, and without these corrections a quantitative fit of the data within the framework of the P.S.-model is impossible.

\newpage

\section{Estimate of the range of exponents compatible with the data}\label{kap5}

In analyzing experimental melting curves usually exponents $\left(c, c_1\right)$ $= \left(1.75, 0 \right)$ are chosen. The network scenario relates $c, c_1$ to exponents governing the partition function of three-armed star polymers.  With the most recent estimates \cite{Z15} for these exponents one finds $\left(c, c_1\right) = (2.15, 0.11).$ We here first examine, whether these two sets of exponents allow for a reasonable fit.

In Fig.~9 we show the melting curves for chains of length $N = 500,$ $3000.$ Clearly, the network based exponents (2.15, 0.11) yield a fit of the same quality (full lines) as found in the previous section. The choice (1.75, 0), (long dashes), however, fails to capture the chain length dependence of the data. Indeed, with an appropriate choice of $a_\tau,$ $a_1,$ $\epsilon^*$ each individual melting curve can be reproduced reasonably well. We here have determined the parameters by fitting $\rho_{bu} \left(N = 3000 \right).$ But the parameters extracted definitely depend on $N.$ This eliminates the choice $c = 1.75,$ $c_1 = 0.$ Further analysis shows that this failure is not related to the value of $c$ but to $c_1 = 0.$ The short dashed curves in Fig.~9 are calculated with exponents $(1.75,$ $0.11),$ which again allow for a reasonable fit.

Keeping $c_1 = 0.11$ fixed but varying $c$ we in the range $1.7 \alt c \alt 2.3$ have found acceptable fits for all quantities related to the distribution of the number of bound pairs. The change in $c$ essentially is compensated by a change of the cooperativity parameter $\sigma,$ which varies from $\sigma\left(c = 1.7\right) \approx 0.03$ to $\sigma\left(c = 2.3\right) \approx 0.45.$ This observation is completely consistent with the result of Ref.~\cite{Z11}. The question to which extent the data for the loop length distribution restrict the value of $c$ will be discussed below.

We now first consider the range of $c_1.$ Fig.~10 shows the distribution of the number of bound pairs for $\epsilon = 1.34264,$ $N = 500.$ The curves give theoretical results for $c = 2.15,$ $0.07 \le c_1 \le 0.20.$ As discussed earlier, we do not put too much weight on the range of small $n \left(n \alt 50, \mbox{i.~e.} \: n/N \alt 0.1 \right),$ where we expect subleading corrections to become relevant. Fig.~10 suggests that $c_1$ should be chosen in the range $0.10 \alt c_1 \alt 0.15.$ This is consistent with all other data. In particular, the height of the maxima in the melting curves is quite sensitive to $c_1.$ (Recall the discussion in the context of Fig.~9.) It increases with increasing $c_1,$ and reasonable fits need a value somewhere between $0.07$ and $0.15.$ Fig.~3a suggests an explanation of this behavior: increasing $c_1$ puts stronger weight on essentially closed or essentially open configurations and makes the transition sharper. We finally note that the estimate $0.10 \alt c_1 \alt 0.15$ is quite independent of the value of $c$ chosen.

We now turn to the determination of $c.$ The estimate $1.7 \alt c \alt 2.3$ quoted above is based on fits of the melting curves. For smaller $N$ the tail to the right of the maximum with increasing $c$ slowly changes from undershooting the data to overshooting. Fig.~9b shows an indication of this effect. The other quantities related to the density of bound pairs are fairly insensitive to $c,$ except that increasing $c$ suppresses the initial peak in ${\cal P}^{[BP]}_{bu}\left(n,N\right)$ for $n \alt 50,$ i.e. in the region where $1/n$ corrections should become relevant.

The loop length distribution might be expected to be more sensitive to $c.$ Fig.~11 shows $\log_{10}\left(\ell^{2.05}{\cal P}^{[LL]}_{bu}\left(\ell , N\right)\right)$ as function of $\log_{10} \ell ,$ where in the theoretical curves we again replaced the explicit factor ${\cal U} \left(\ell\right)$ by $\sigma \ell^{-c},$ cf. Eq.~(2.60). We omitted the region $\log_{10} \ell > 2.5,$ where the theoretical curves for all $c$ in the range $1.7 \alt c \alt 2.3$ essentially coincide. We also omitted the data for $N = 1280,$ since the accuracy rapidly decreases for $\log_{10} \ell > 2,$ where these data deviate from those shown. Besides the curves for $\left(c, c_1\right) = (2.05, 0.13),$ (full lines, cf. Fig.~7), we included the results for (1.90, 0.11), (long dashes), and (2.20, 0.11), (short dashes). For given $N$ the curves merge for $\ell \agt 10^2.$ For smaller $\ell$ the theoretical curves as function of $c$ sweep over the data. 
The behavior is consistent with the parameters $\tilde{\cal U}_s \left(1\right),$ $\tilde{\cal U}^{\prime}_s \left(1\right)$ extracted, which show that for $c = 1.90$ the here neglected part ${\cal U}_s \left(\ell\right)$ is predominantly positive, whereas it is predominantly negative for $c = 2.20.$ We furthermore recall from the general discussion of Sect.~\ref{kap2} that the first moment $\left<\ell\right>_{bu}$ of ${\cal P}^{[LL]}_{bu}$ can be expressed in terms of $\rho_{bu},$ $\left<\bar{m}\right>_{bu},$ and $1-w^*,$ cf.~Eq.~(2.22). All the parameter sets used here essentially yield the same results for these quantities and thus for $\left<\ell\right>_{bu}.$ This suggests that for each parameter set we could find a short-range part ${\cal U}_s \left(\ell\right)$ of ${\cal U}\left(\ell\right),$ so that the data for ${\cal P}^{[LL]}_{bu}$ are fitted consistently together with all other quantities. An example is shown in the insert of Fig.~11 and will be discussed below. In view of the above, the problem of fixing a range of $c$ from data for ${\cal P}^{[LL]}_{bu}$ amounts to identifying a value $\ell_0$ such that ${\cal U}_{s} \left(\ell\right) \ll \ell^{-c}$ for $\ell > \ell_0 .$ For estimating $\ell_0$ the literature offers two somewhat contradictory pieces of evidence. 
Ref.~\cite{Z3} presents results of exact enumeration on the square lattice, for chain lengths $N \le 15.$ The data seem to obey the scaling law ${\cal P}^{[LL]}_{bu} \left(\ell, N\right) = \ell^{-c}$ $\hat{\cal P} \left(\ell / N\right)$ with $c = 2.44 \pm .06.$ The network scenario predicts $c \approx 2.41$ in two dimensions. These results suggest $\ell_0 \approx 10,$ and accepting the same value in three dimensions we would conclude that $c$ is very close to $c = 2.05.$ The network scenario predicts $c = 2.15,$ an estimate based on simulations of star polymers \cite{Z15} with arm lengths up to $n = 4000.$ However, previous work \cite{Z16} covering only arm lengths $n \alt 130$ resulted in somewhat different effective exponents, which yield a value $c \approx 2.10.$ Thus, in an optimistic view we might take the good fit with $c = 2.05$ as supporting the network scenario. This value could be interpreted as an effective exponent describing loop lengths of order $\ell \approx 10^2.$ Consistent with the above mentioned results the effective exponent is smaller than the asymptotic value $c \approx 2.15.$ Clearly this interpretation implies that ${\cal U}_s \left(\ell\right)$ is a sizeable correction up to at least $\ell_0 \approx 10^2.$ This is quite plausible, since the corrections to asymptotic scaling in the excluded volume problem in three dimensions are known to decrease roughly like $\ell^{-0.5}.$ The insert in Fig.~11 demonstrates that such corrections easily can bring the theoretical curves to match the data also for $c \neq 2.05.$ (In two dimensions the corrections are expected to decrease like $1/N$, which might explain the observation of Ref.~\cite{Z3} quoted above.)

However, irrespective of the validity of any specific choice of $c$ these considerations imply that ${\cal U}_s\left(\ell\right)$ cannot be expected to be negligible for $\ell \alt 10^2.$ Whatever the true value of the exponent $c$ might be, we must expect that embedding of the chain into three-dimensional space gives rise to corrections typical of the excluded volume problem. In a less optimistic view we thus would conclude that $c$ might take any value in about the range $1.9 \alt c \alt 2.2,$ these values resulting from the assumption that $\ell_0$ is of the order of 100. Even though the choice $c = 2.15$ based on the network scenario allows for a consistent interpretation of the data, we from the analysis of the data have no reliable arguments to exclude other values of $c.$

\newpage

\section{Analysis of other observables}\label{kap6}
According to our analysis the present data are compatible with exponents $c$ in at least the fairly large range $1.9 \alt c \alt 2.2.$ For a more precise determination the obvious approach would be to simulate longer chains. For instance,  with chains of length $N = 30000$ the predicted height of the maximum of the melting curve increases by about $25\%$ in going from $c = 1.9$ to $c = 2.2.$ However, it seems unlikely that for such long chains sufficiently precise simulations can be carried through in the near future. We thus should look for other observables sensitive to $c.$ 

Two features of the transition prevent a precise determination of $c.$ Firstly, the smallness of $|c-2|$ forces us to include the corrections to finite size scaling with the associated fit parameter $a_1.$ Only the simulation of much longer chains could eliminate these terms, which blur the qualitative difference among $c <2$ or $c > 2.$ Secondly, with the $(bu)$-boundary conditions used the chain essentially opens from the end, irrespective of $c.$ For instance, for $\bar\tau = 0$ on average about half of the chain is open for all $c$ in the above range, irrespective of $N.$ Clearly, this feature suppresses the sensitivity to $c$ of the transition. To eliminate this effect we should switch to $(bb)$-boundary conditions, where both chain ends are closed.

Fig.~12 shows the density of bound pairs and the melting curves predicted with $(bb)$-boundary conditions for $N = 500,$ $\left(c, c_1\right) = \left(2.20,\: 0.11 \right)$ or $\left(1.90, \: 0.11\right),$ respectively. The parameters $\epsilon^*, a_\tau, a_1$ are taken from the fit to the $(bu)$-data. We note that the transition is shifted towards negative values of $\left(\epsilon - \epsilon^*\right) N,$ as expected. What is more important, even for such short chains we see a clear effect of changing $c.$ For $\rho_{bb}$ the effect just increases with $N.$ For $C_{bb}/N$ the peak height for $c = 1.90$ decreases faster than for $c = 2.20.$ Near $N = 1000$ there is a region where the peak heights approximately coincide and where an experimental discrimination among the two predictions will be more difficult.

In Fig.~13 we compare predictions for the mean length of the first loop. Panel $a)$ shows results for $(bu)$-boundary conditions. For $N = 500$ the effect of changing $c$ is quite small, but it increases rapidly with $N.$ Data for $N = 3000$ could improve the estimate of $c$ considerably. As panel $b)$ shows, for $(bb)$-boundary conditions the effect again is increased, to the level where data for $N = 500$ could be as useful as data for $N = 3000$ and $(bu)$-boundary conditions.

These findings suggest that a more precise determination of the exponent $c$ would be possible by comparing the results for $\left(bu\right)$- or $(bb)$-boundary conditions. Such simulations could be carried through for fairly short chains, since a clear effect is predicted even for $N = 500.$

\newpage

\section{Conclusions}\label{kap7}

We have found that a model of the type proposed by Poland and Scheraga is able to quantitatively reproduce simulation data of a very simple model of the DNA denaturation transition. It, however, turned out to be essential to amend the original P.S.-model by a factor $p\left(m, N\right)$ weighting the unbound ends of the doubly stranded chain. The simple ansatz $p\left(m, N\right) = m^{-c_1} \left(N - m \right)^{-c_1}$ is sufficient, but without such a factor the model cannot reproduce the chain length dependence of the data. Concerning the application of the P.S.-model to the analysis of physical melting curves, this may be the most important result found here.

Concerning the simulation data we have found that an interpretation within the framework of the P.S.-model implies that chain lengths $N \alt 3000$ are far too short for asymptotic finite size scaling to hold. We must include corrections to this limit which decay only as $N^{-|c-2|}.$ Since the numerical value of $|c-2|$ is quite small, this implies that all the effective exponents extracted by straight forward data analysis are strongly influenced by the correction terms. In particular this holds for the exponents which motivated the co-polymer network scenario. For the dangling ends this scenario introduces a weight $p\left(m, N\right)$ involving two different exponents. Our analysis lends no support to this hypothesis.

Allowing for unbinding of an end of the doubly stranded chain we have found that denaturation is dominated by the prolongation of the dangling ends. The internal loop structure of the closed part of the chain does not change much during the transition. As pointed out in Sect.~\ref{kap3}, for $c > 2$ this is true irrespective of $N,$ but for the chain lengths considered here it holds true also for $c < 2.$ We, for instance, note that for $\left(c, c_1\right) = \left(1.75, 0.11 \right),$ $N = 3000,$ the mean length of a loop according to the theory increases only from $\left<\ell\right>_{bu} \approx 4.7$ to $\left<\ell\right>_{bu}\approx 7.1$ in a range where the average length of the open ends increases from $\left<m\right> = 0.1 \: N$ to $\left<m \right> = 0.9 \: N.$ This observation is of immediate consequences, if we try to determine the exponents $\left(c, c_1\right)$ from the data. $c_1,$ which governs the weight of the dangling ends, can be determined with fair precision. We found $0.10 \alt c_1 \alt 0.15.$ The loop exponent $c$, however, at best is bounded by $1.9 \alt c \alt 2.2.$ An even larger range results if we allow for more pronounced short range effects in the ansatz for the loop weight ${\cal U} \left(\ell\right).$ In full agreement with previous work \cite{Z11} we find that changes of $c$ essentially are compensated by changing the cooperativity parameter $\sigma.$

Excluding simulation of much longer chains we see two possibilities to decrease the uncertainty of $c.$ We found that the mean length $\left<\ell_1 \right>$ of the first loop is sensitive to $c,$ even if we allow the chain end to unbind. For $N \approx 3000$ the effect should be measurable reasonably well. Another, and possibly more efficient, approach would be to simulate chains bound together at both ends. This eliminates the pathway dominating denaturation for unbound ends. The theory predicts measurable effects of changing $c$ even for $N \approx 500.$ Furthermore such boundary conditions have the additional virtue of suppressing the factor $p\left(m, N\right)$, which is not well known quantitatively.

In summary, the results of the present work suggest that the Poland-Scheraga model generalized by an end-weighting factor catches the essential physics of the denaturation transition. Since the network scenario predicts such a factor we feel that the analysis also supports this scenario with its associated exponents $\left(c, c_1\right) = \left(2.15, 0.11\right).$ However, more work must be done to put this latter conclusion on a firmer basis.


\nop
\begin{eqnarray}
1 + 1 = 2
\end{eqnarray}

\begin{equation}
1 + 1 = 2
\end{equation}

\mathop{\,\hbox to 2.5em{$-$\hskip-1ex\leaders\hbox to1ex{\hss$-$\hss}\hfill
        \hskip-1ex$\longrightarrow$}\,}\limits_{N\to\infty}

\begin{eqnarray}
{\cal Z}_{bb}\left(N\right) = \sum^{\infty}_{k = 0} \:\: & &\sum_{\{j_0, j_1, \cdots ,j_k\} \ge 1 \atop \{\ell_1, \cdots ,\ell_k\} \ge 1} \delta \Big(N - j_0 - \sum^{K}_{\mu = 1}\left(j_\mu + \ell_\mu\right)\Big) w^{j_0} \prod^{K}_{\mu = 1} \left({\cal U}(\ell_\mu) w^{j_\mu}\right)~. 
\end{eqnarray}
\nopend


\newpage 

\begin{figure}
\label{fig1}
\begin{center}
\epsfig{figure=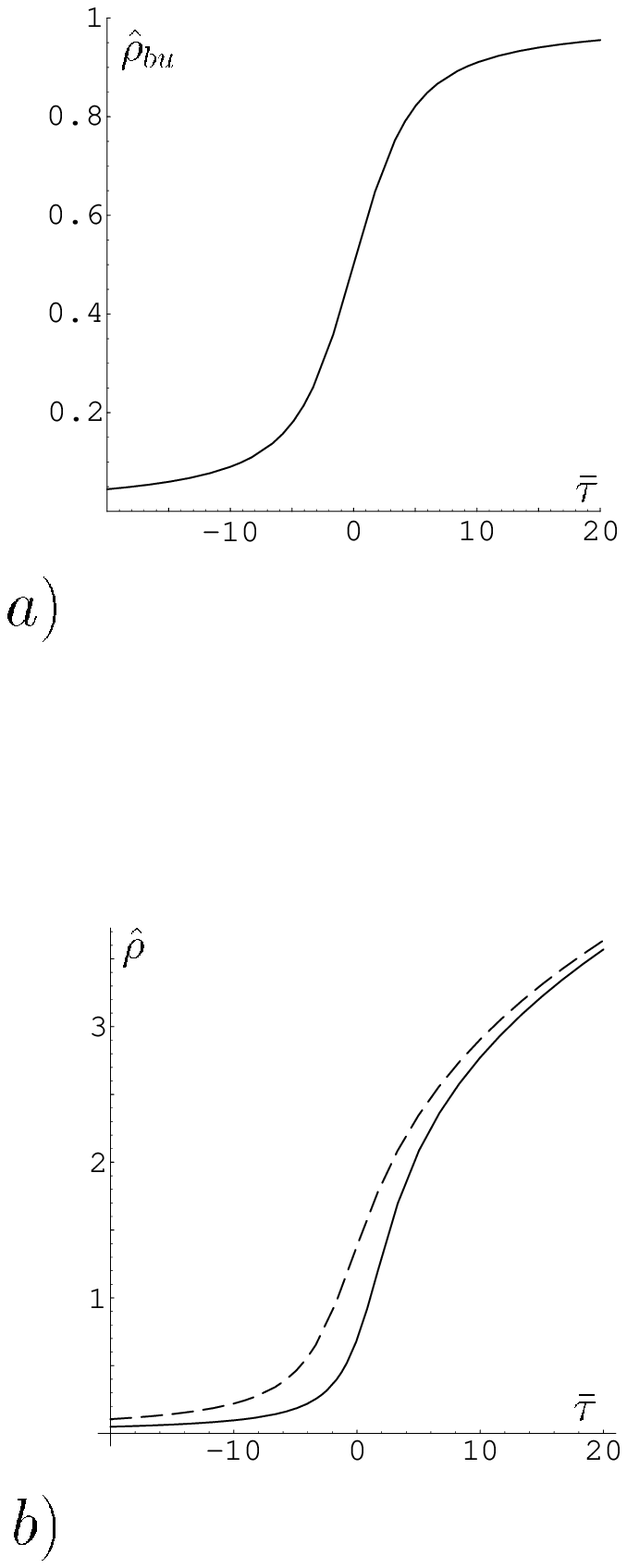,width=0.5\linewidth} 
\caption[]{Scaling function $\hat\rho$ of the density of bound pairs as function of $\bar\tau$. $a)$ $\hat\rho_{bu}$ for $c_1 = 0.11,$ $c >2.$ $b)$ $\hat\rho_{bu}$ (full line) and $\hat\rho_{bb}$ (broken line) for $c_1 = 0,$ $c = 1.75.$}
\end{center}
\end{figure}

\begin{figure}
\label{fig2}
\begin{center}
\epsfig{figure=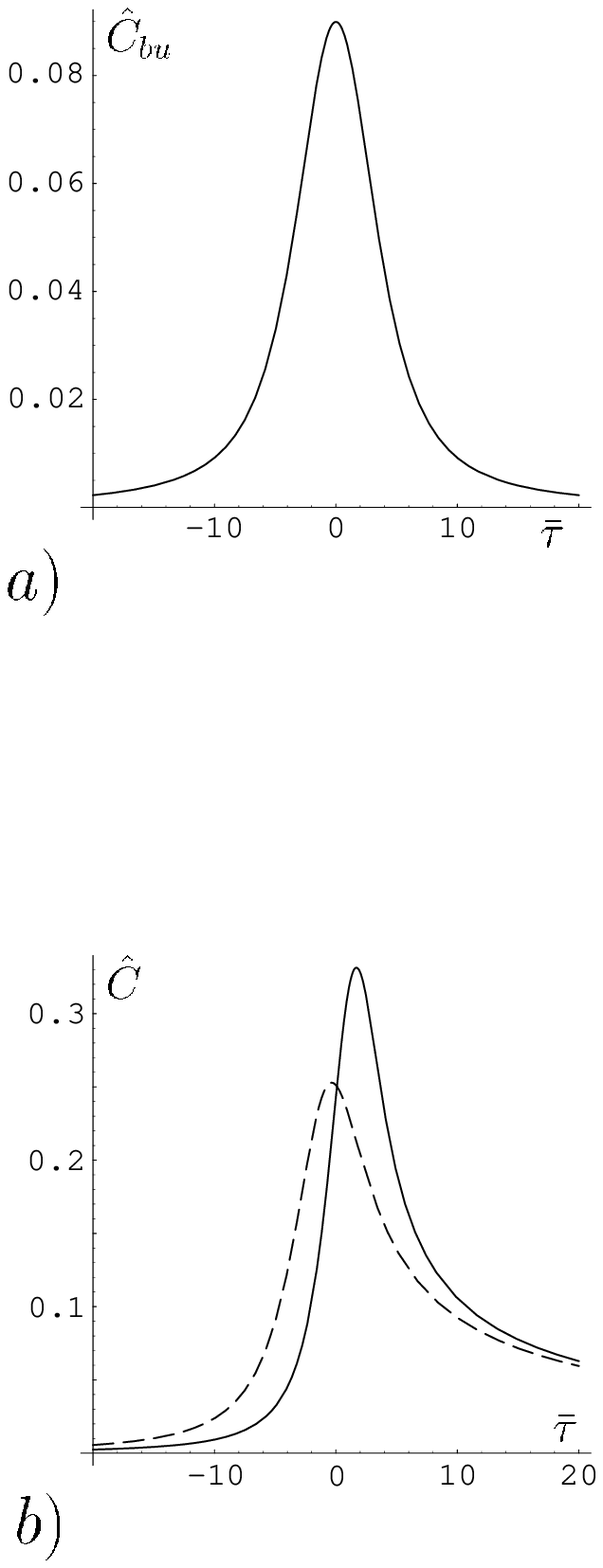,width=0.5\linewidth} 
\caption[]{Scaling functions $\hat{C}$ of the fluctuations in the number of bound pairs, (melting curves). $a)$ $\hat{C}_{bu}$ for $c_1 = 0.11,$ $c> 2.$ $b)$ $\hat{C}_{bu}$ (full line) and $\hat{C}_{bb}$ (broken line) for $c_1 = 0,$ $c = 1.75.$}
\end{center}
\end{figure}

\begin{figure}
\label{fig3}
\begin{center}
\epsfig{figure=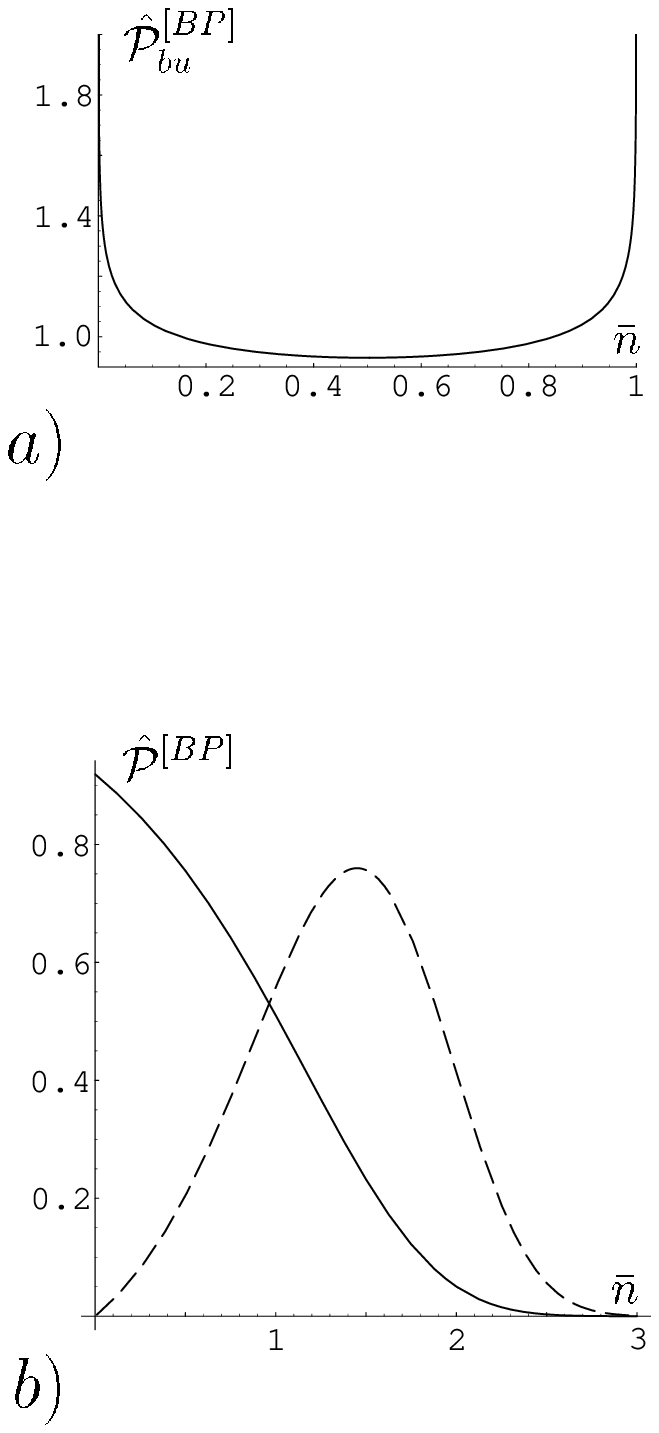,width=0.5\linewidth} 
\caption[]{Scaling functions $\hat{\cal P}^{[BP]}$ of the distribution of the number of bound pairs as function of $\bar{n}$ at the critical point $\bar\tau = 0.$ $a)$ $\hat{\cal P}^{[BP]}_{bu}$ for $c_1 = 0.11,$ $c >2.$ $b)$ $\hat{\cal P}^{[BP]}_{bu}$ (full line) and $\hat{\cal P}^{[BP]}_{bb}$ (broken line) for $c_1 = 0,$ $c = 1.75.$}
\end{center}
\end{figure}

\begin{figure}
\label{fig4}
\begin{center}
\epsfig{figure=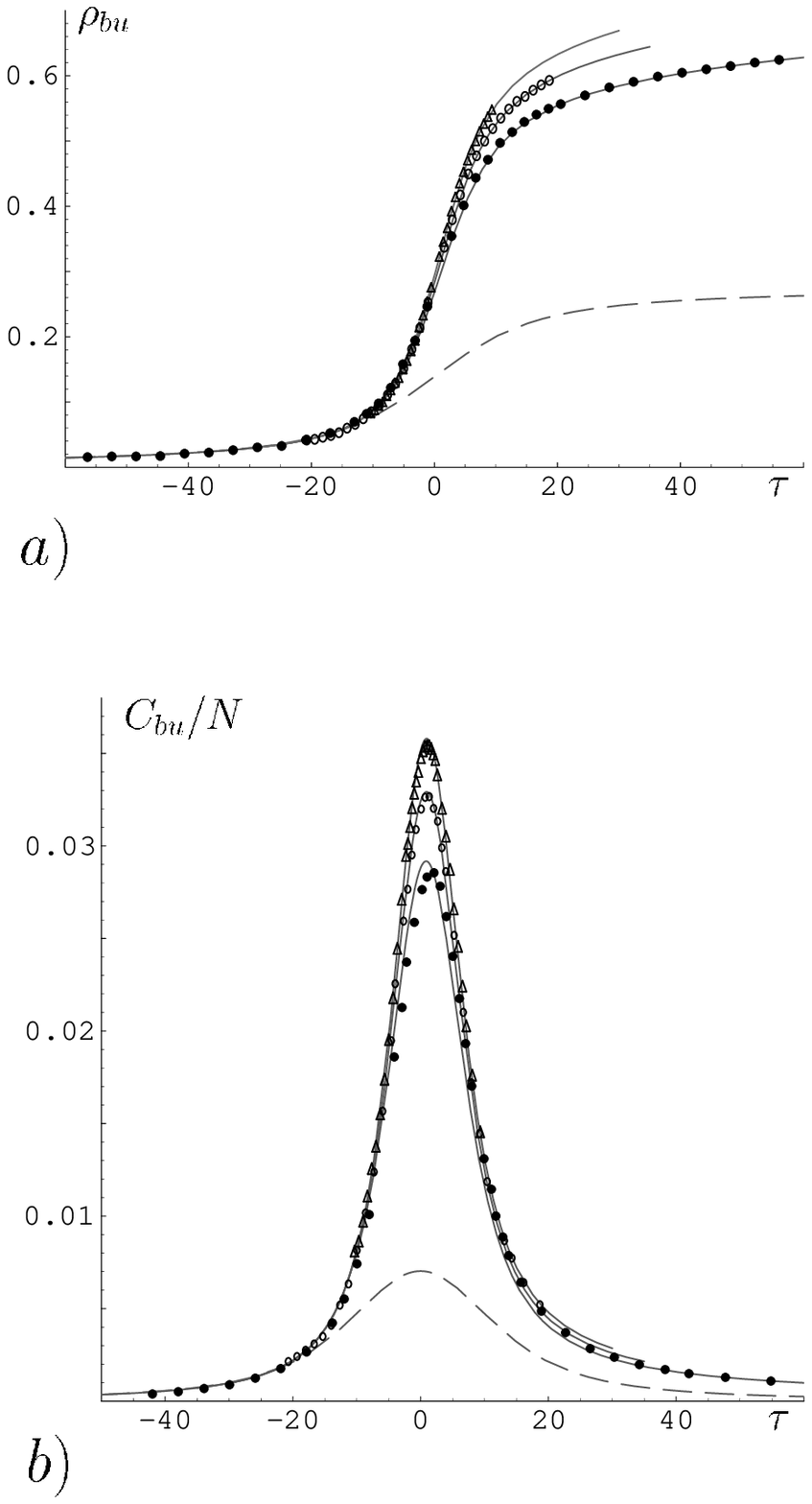,width=0.5\linewidth} 
\caption[]{Bond density $\rho_{bu}$ (panel $a)$) and melting curves $C_{bu}/N$ (panel $b)$) as function of $\tau = \left(\epsilon - \epsilon^* \right)N.$ Data from Ref.~\cite{Z1}. Triangles: $N = 500;$ circles: $N = 1000;$ points: $N = 3000.$ Curves: P.S.-model with $(c,c_1) = (2.05,$ $0.13)$. The broken curves give the finite size scaling limit $N \to \infty.$}
\end{center}
\end{figure}

\begin{figure}
\label{fig5}
\begin{center}
\epsfig{figure=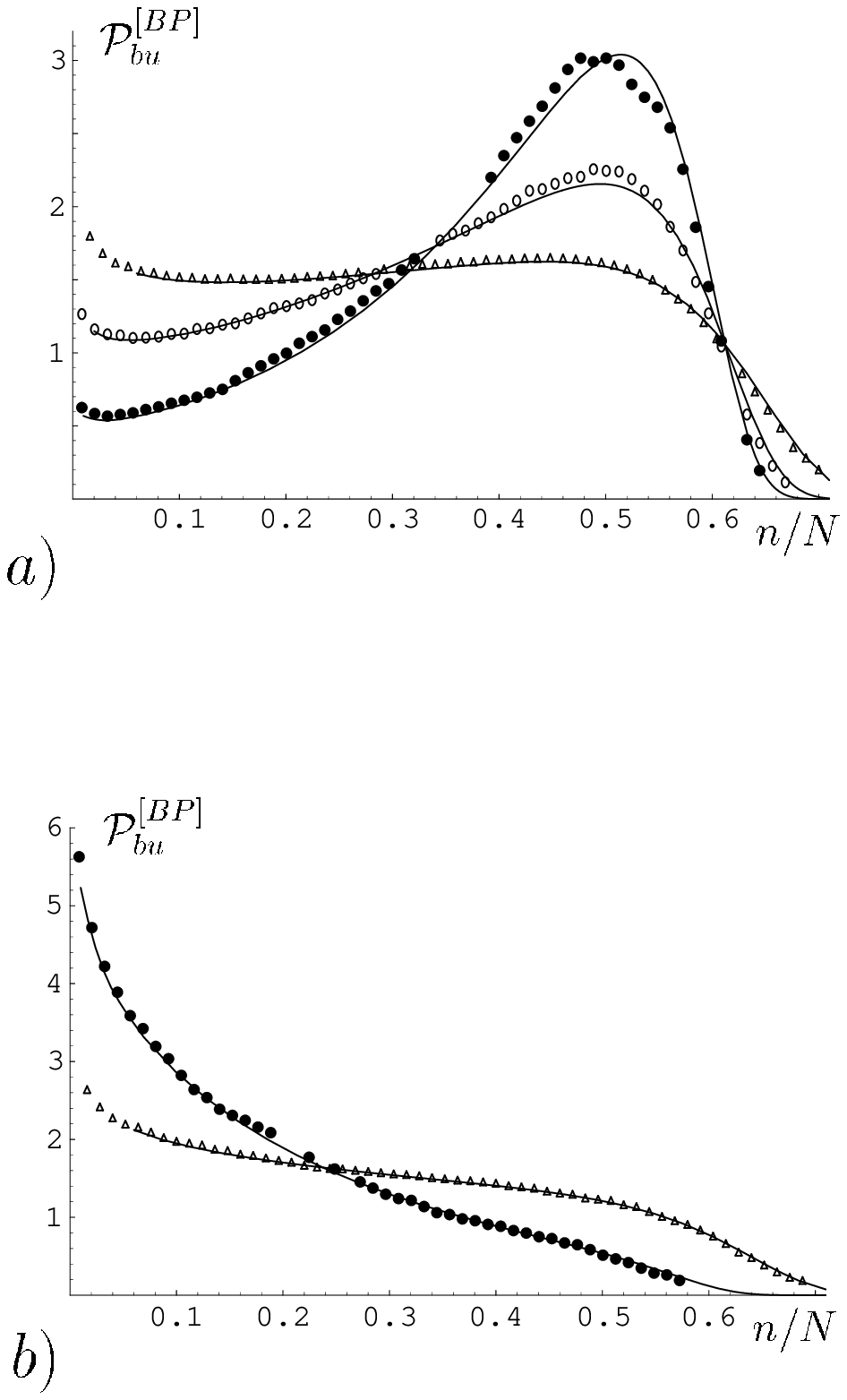,width=0.5\linewidth} 
\caption[]{
Distribution function ${\cal P}^{[BP]}_{bu}$ of the number $n$ of bound pairs as function of $n/N.$ Data from Ref.~\cite{Z1}. Triangles: $N = 500;$ circles: $N = 1500;$ points: $N = 3000.$ Panel $a)$: $\epsilon = 1.34264 > \epsilon^*;$ panel $b)$: $\epsilon = 1.33996 < \epsilon^*.$ Curves are the results of the P.S.-model.
}
\end{center}
\end{figure}

\begin{figure}
\label{fig6}
\begin{center}
\epsfig{figure=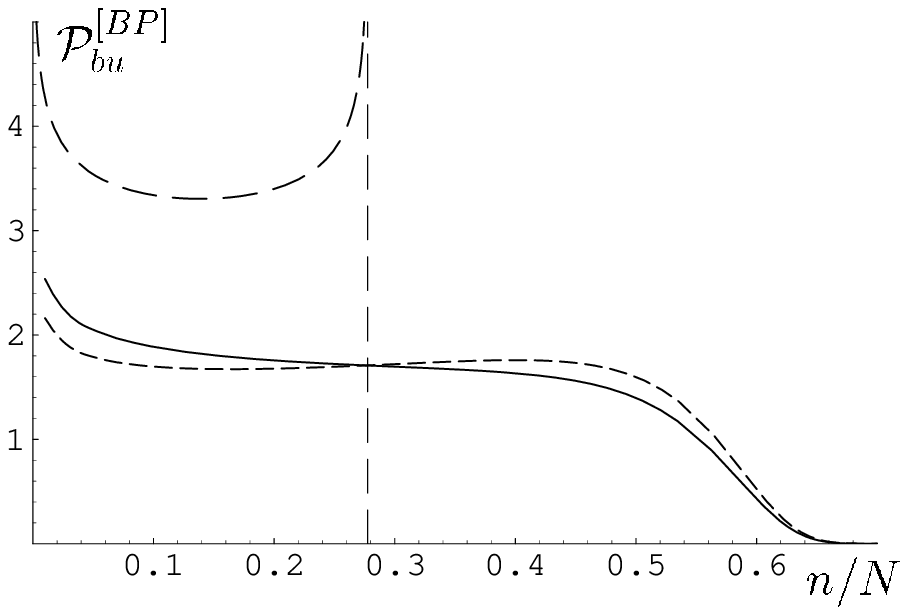,width=0.5\linewidth} 
\caption[]{
${\cal P}^{[BP]}_{bu}$ as predicted by the P.S.-model. See the main text for a discussion.
}
\end{center}
\end{figure}

\begin{figure}
\label{fig7}
\begin{center}
\epsfig{figure=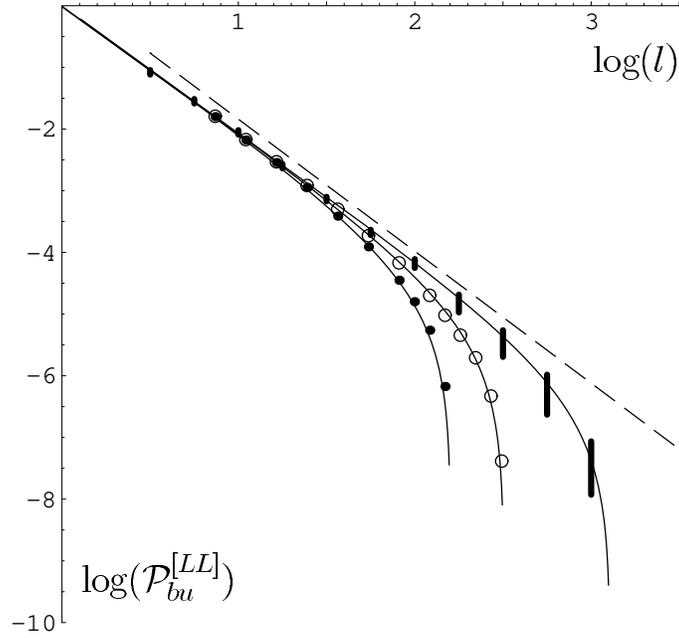,width=0.5\linewidth} 
\caption[]{
$\log_{10} {\cal P}^{[LL]}_{bu}$ as function of $\log_{10} \ell.$ Data from Refs.~\cite{Z3,Z4}. Points: $N = 160;$ circles: $N = 320;$ bars: $N = 1280.$ Curves are calculated from Eqs.~(2.60), (2.61), with $\cal U \left(\ell \right)$ replaced by $0.14$ $\ell^{-2.05}.$ The broken line represents the effective power law $\ell^{-2.14}.$
}
\end{center}
\end{figure}

\begin{figure}
\label{fig8}
\begin{center}
\epsfig{figure=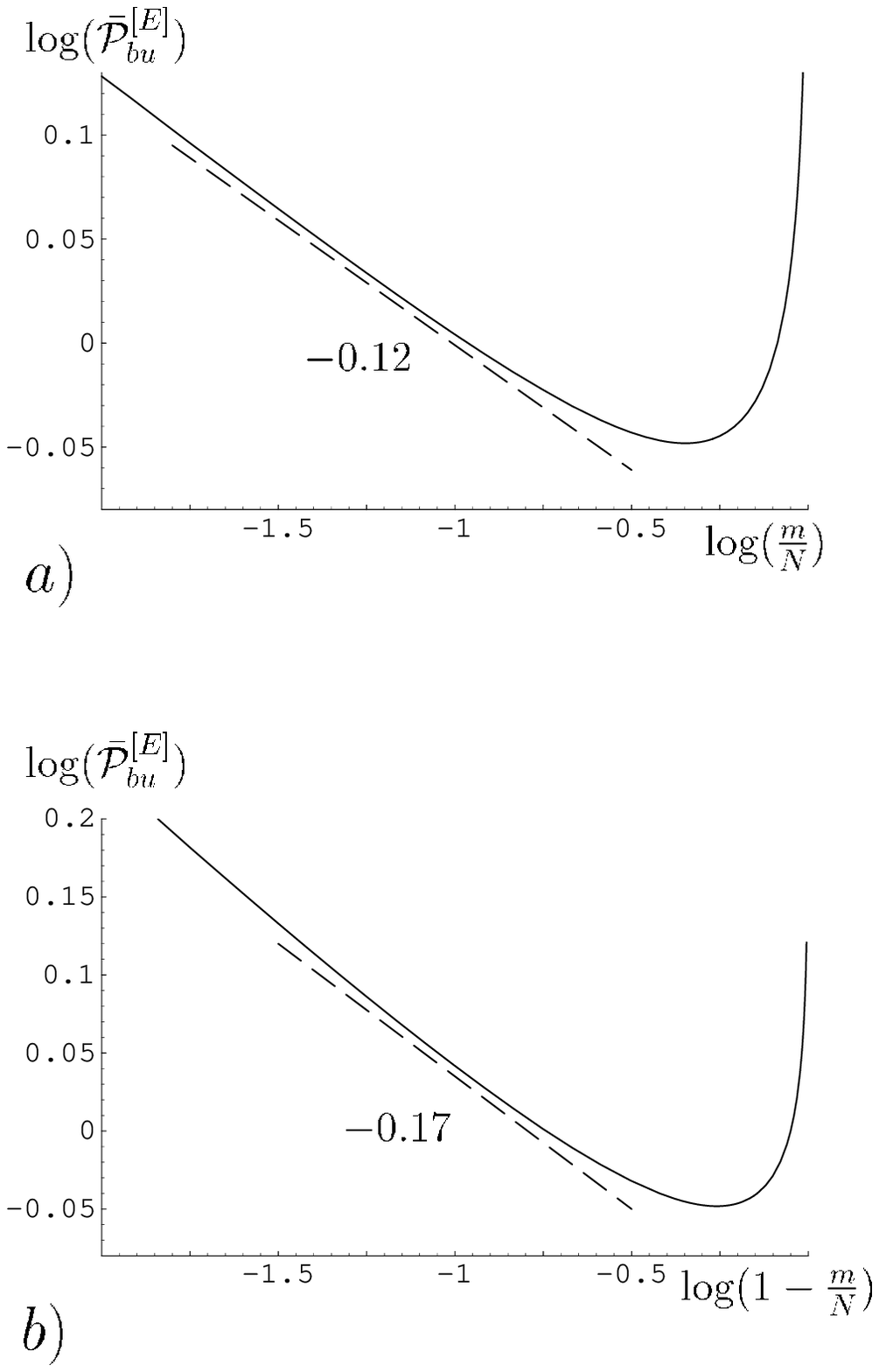,width=0.5\linewidth} 
\caption[]{
$\log_{10} \bar{\cal P}^{[E]}_{bu}$ as function of $\log_{10} \left(m/N\right)$, panel $a)$, or $\log_{10} \left(1 - m/N \right)$, panel $b)$, for $N = 320,$ $\epsilon = 1.3413.$ The broken lines indicate slopes $-0.12,$ (panel $a)$), or $-0,17,$ (panel $b)$), respectively.
}
\end{center}
\end{figure}

\begin{figure}
\label{fig9}
\begin{center}
\epsfig{figure=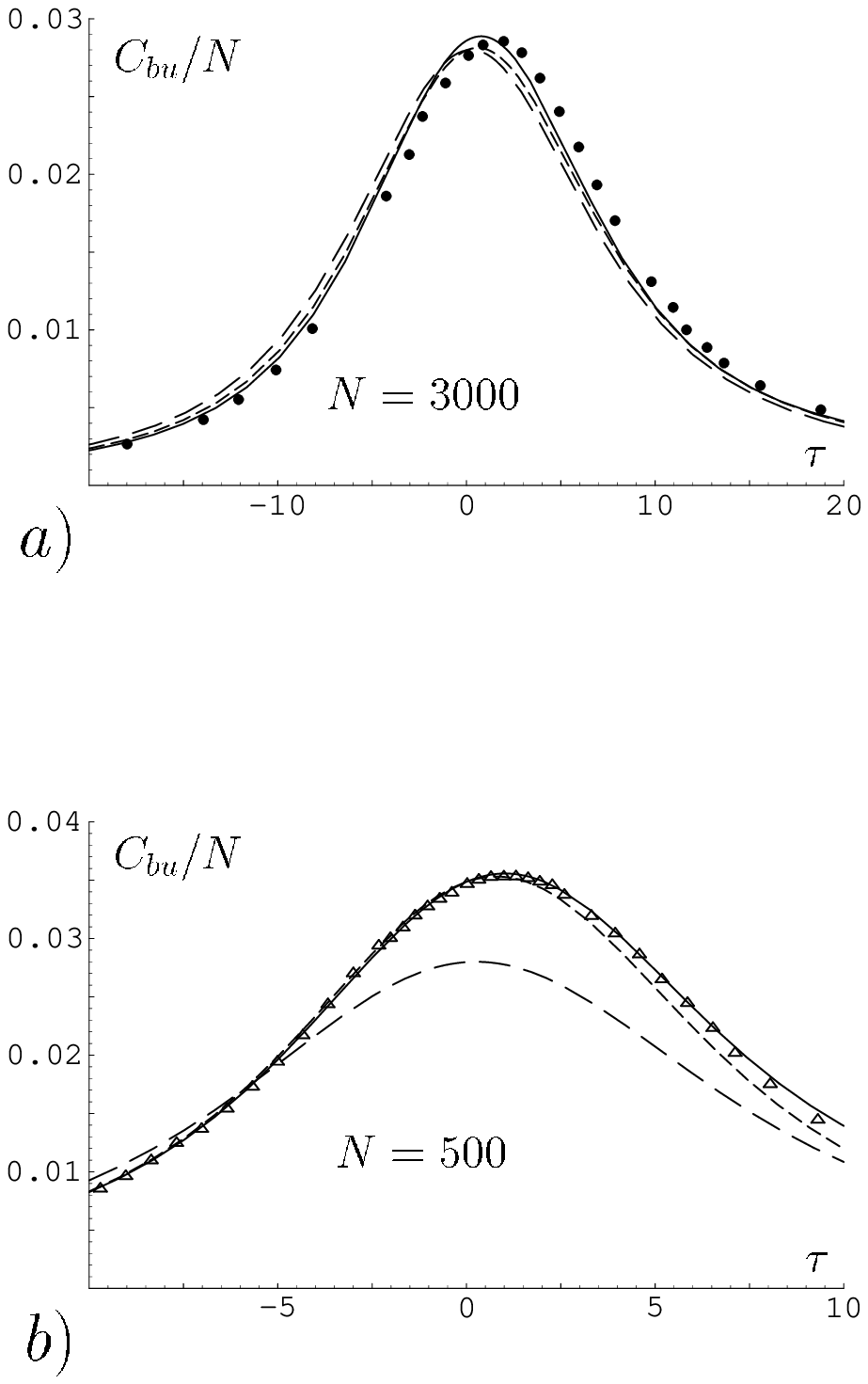,width=0.5\linewidth} 
\caption[]{
Melting curves $C^{[bu]}/N$ as function of $\tau = N \left(\epsilon - 1.34114\right)$, where $\epsilon^* = 1.34114$ has been determined with $\left(c, c_1 \right) = \left(2.15, \: 0.11\right)$. Panel $a)$: $N = 3000,$ panel $b)$: $N = 500.$ Data are taken from Ref.~\cite{Z1}. The curves give the results of the P.S.-model for exponents $(2.15,$ $0.11)$, full lines; $(1.75, $ $0)$, long dashes; $(1.75,$ $0.11)$, short dashes.
}
\end{center}
\end{figure}

\begin{figure}
\label{fig10}
\begin{center}
\epsfig{figure=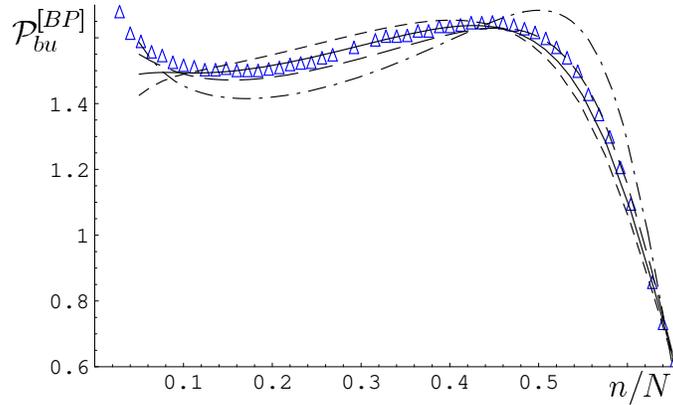,width=0.5\linewidth} 
\caption[]{
${\cal P}^{[BP]}_{bu} \left(n/N \right)$ for $N = 500,$ $\epsilon = 1.34264.$ Data from Ref.~\cite{Z1}. The theoretical curves are calculated for $c = 2.15$ and values $c_1 = 0.07,$ short dashes; $0.11,$ full line; $0.15,$ long dashes; $0.20,$ dot-dashed curve. 
}
\end{center}
\end{figure}

\begin{figure}
\label{fig11}
\begin{center}
\epsfig{figure=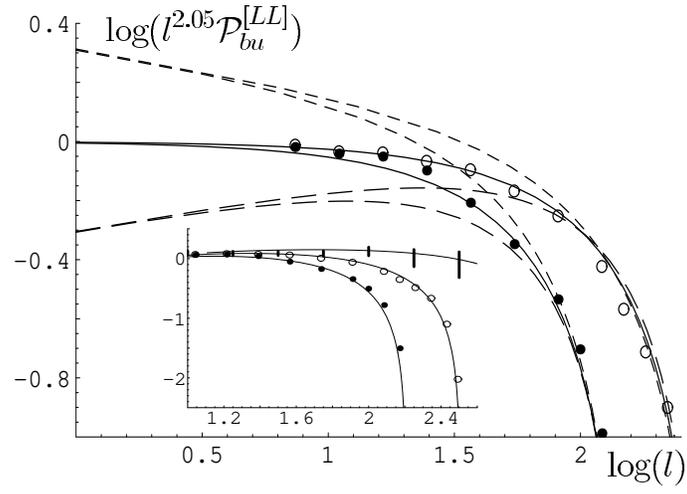,width=0.5\linewidth} 
\caption[]{
$\log_{10} \left(\ell^{2.05} {\cal P}^{[LL]}_{bu}\right)$ as function of $\log_{10} \ell.$ Data taken from Ref.\cite{Z3}. Circles: $N = 320;$ points: $N = 160.$ Theoretical curves are shown for $(c, c_1) = (2.05,$ $0.13 )$, full line: $(1.90,$ $0.11)$, long dashes; $(2.20,$ $0.11)$, short dashes. The insert shows the results for $(c, c_1) = (2.15,$ $0.11)$ and ${\cal U} \left(\ell\right) = 0.236$ $\ell^{-2.15} \left(1 - \ell^{-0.50} \right)$.
}
\end{center}
\end{figure}

\begin{figure}
\label{fig12}
\begin{center}
\epsfig{figure=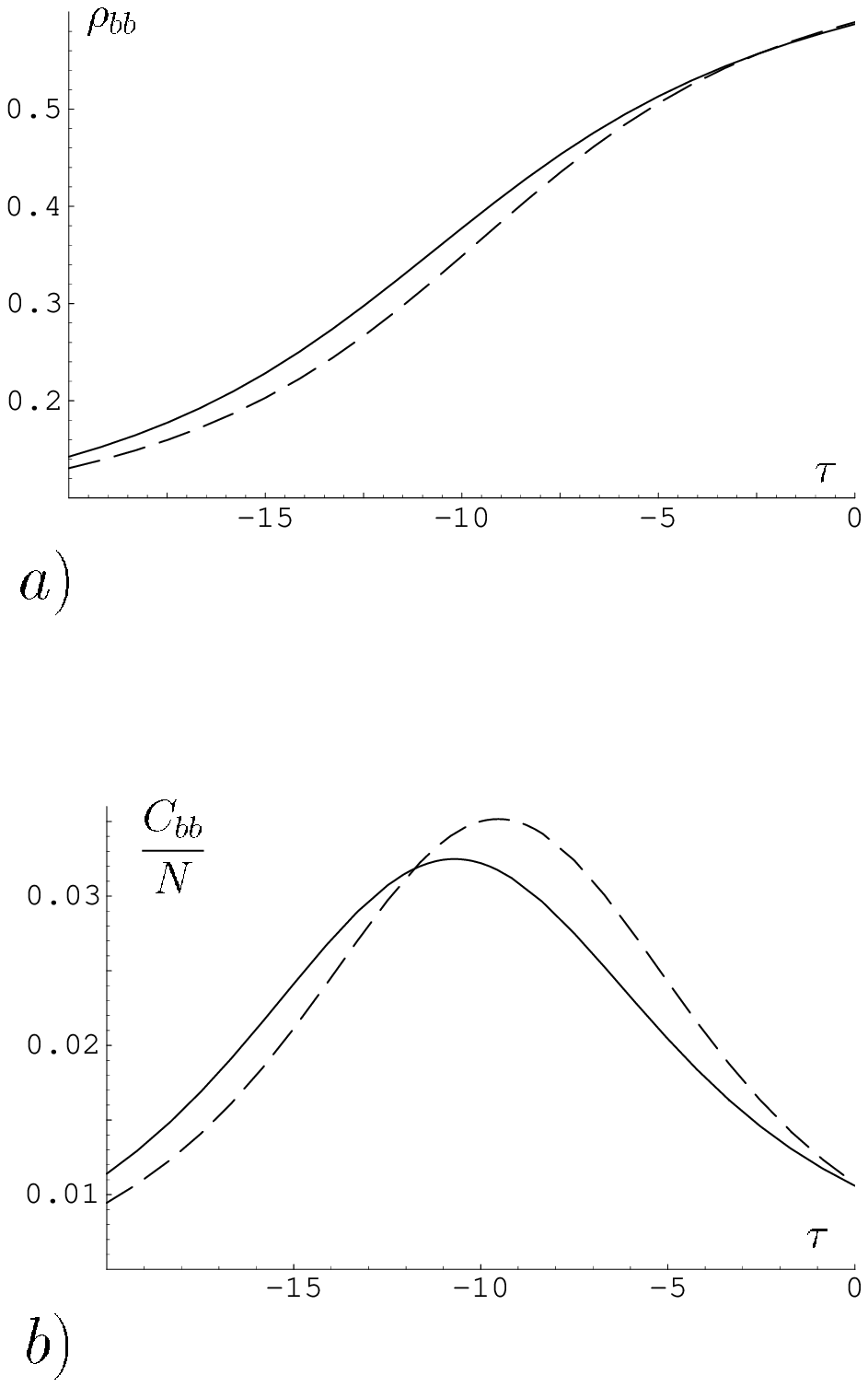,width=0.5\linewidth} 
\caption[]{
$\rho_{bb},$ (panel $a)$), and $C_{bb}/N,$ (panel $b)$), as function of $\tau = \left(\epsilon - 1.34116 \right) N$ for $N = 500.$ Full lines: $(c, c_1) = (2.20,$ $0.11);$ broken lines: $(1.90,$ $0.11)$. In calculating $\tau$ we use  $\epsilon^* = 1.34116$ as determined for $(c, c_1) = (2.20,$ $0.11)$.
}
\end{center}
\end{figure}

\begin{figure}
\label{fig13}
\begin{center}
\epsfig{figure=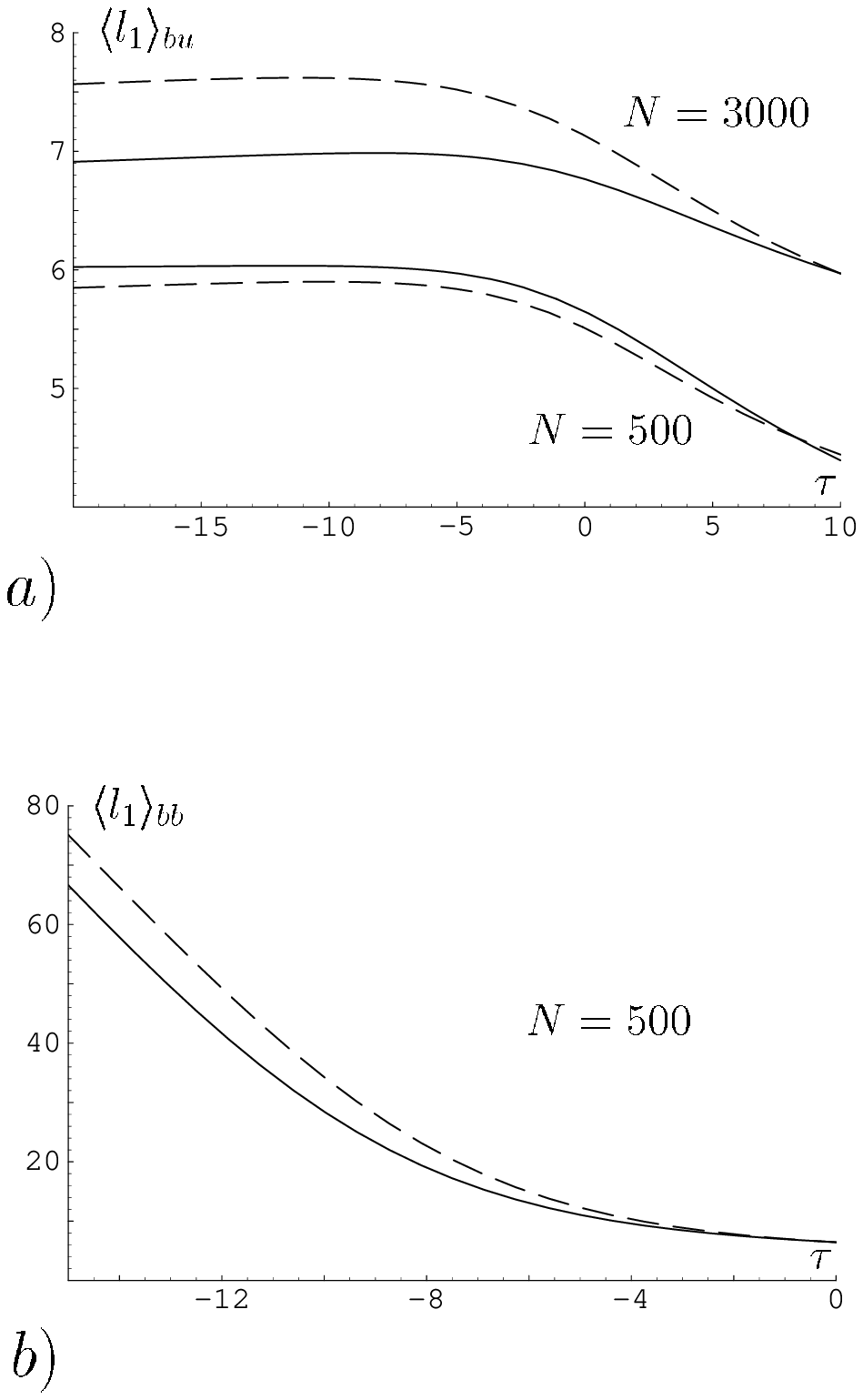,width=0.5\linewidth} 
\caption[]{
Mean length of the first loop calculated for $(c, c_1) =$ $(2.20,$ $0.11)$, full lines, or $(1.90,$ $0.11)$, broken lines and different boundary conditions. $a)$ $bu$-conditions; $b)$ $bb$-conditions. $\tau = (\epsilon -  1.34116) N$; chain lengths as given in the figures.
}
\end{center}
\end{figure}


\begin{thebibliography}{88}
\bibitem{Z1} M.S.~Causo, B.~Coluzzi, P.~Grassberger, Phys. Rev. E {\bf 62},
		 3958 (2000).
\bibitem{Z2} E. Carlon, E. Orlandini, A.L. Stella, Phys. Rev. Lett. {\bf 88}, 
		198101 (2002).
\bibitem{Z3} M. Baiesi, E. Carlon, A.L. Stella, Phys. Rev. E {\bf 66}, 021804 				
		(2002).
\bibitem{Z4} M. Baiesi, E. Carlon, Y. Kafri, S. Mukamel, E. Orlandini,
		 A.L.~Stella, Phys. Rev. E {\bf 67}, 021911 (2003).
\bibitem{Z5} D. Poland, H.A. Scheraga, J. Chem. Phys. {\bf 45}, 1456 (1966).                 
\bibitem{Z6} D. Poland, H.A. Scheraga, {\it Theory of Helix Coil Transition in 
		Biopolymers}, (Acad.~Press, New~York, 1970).
\bibitem{Z7} R.M. Wartell, A.S. Benight, Physics Reports {\bf 126}, 67 (1985).
\bibitem{Z8} R.D. Blake, S.G. Delcourt, Nucleic Acids Res. {\bf 26}, 3323 	
		(1998).
\bibitem{Z9} M.E. Fischer, J. Chem. Phys. {\bf 45}, 1469 (1966).
\bibitem{Z10} Y. Kafri, D. Mukamel, L. Peliti, Phys. Rev. Lett. {\bf 85}, 4988 
		(2000).     
\bibitem{Z11} R. Blossey, E. Carlon, Phys. Rev. E {\bf 68}, 061911 (2003).
\bibitem{Z12} C. Richard, A.J. Guttmann, J. Stat. Phys. {\bf 115}, 925 (2004).
\bibitem{Z13} Y. Kafri, D. Mukamel, L. Peliti, Eur. Phys J. B~{\bf 27}, 135 
		(2002).              
\bibitem{Z14} The notion `co-polymer network' is somewhat misleading. It is known that all networks of given topology show the same asymptotic scaling properties, irrespective of chemical composition, provided only that all interactions among the monomers are repulsive. The hypothesis put forward in Ref.~\cite{Z3} is not concerned with a network of given topology. It rather postulates that a polymer chain decorated with a fixed density of loops, the loop lengths being power law distributed, belongs to a new universality class, different from the self-avoiding walk class.
\bibitem{Z15} H.-P. Hsu, W. Nadler, P. Grassberger, Macromolecules {\bf 37}, 
		4658 (2004).
\bibitem{Z16} J. Batoulis, K. Kremer, Macromolecules {\bf 22}, 4277 (1989). 
\end{thebibliography}
\end{document}